\documentclass[lettersize,journal]{IEEEtran}
\usepackage{ifpdf}

\usepackage{amssymb}
\usepackage{color}

\usepackage{cite}

\ifCLASSINFOpdf
  \usepackage[pdftex]{graphicx}
\else
  \usepackage[dvips]{graphicx}
\fi
\usepackage{amsmath}
\usepackage{algorithm}
\usepackage{amsfonts}
\usepackage{algorithmic}

\usepackage{array}

\ifCLASSOPTIONcompsoc
\usepackage[caption=false,font=normalsize,labelfont=sf,textfont=sf]{subfig}
\else
    \usepackage[caption=false,font=footnotesize]{subfig}
\fi
\usepackage{fixltx2e}

\usepackage{stfloats}
\ifCLASSOPTIONcaptionsoff
    \usepackage[nomarkers]{endfloat}
    \let\MYoriglatexcaption\caption
    \renewcommand{\caption}[2][\relax]{\MYoriglatexcaption[#2]{#2}}
\fi
\usepackage{url}

\usepackage{algorithmic}
\begin{document}
%
\title{Energy-Latency Aware Intelligent Reflecting Surface Aided Multi-cell Mobile Edge Computing}
%
%
%

\author{Wenhan~Xu,~\IEEEmembership{Student Member,~IEEE,}
Jiadong~Yu,~\IEEEmembership{Member,~IEEE,}
Yuan~Wu,~\IEEEmembership{Senior Member,~IEEE,}
and~Danny~H.K.~Tsang,~\IEEEmembership{Fellow,~IEEE,}
\thanks{Manuscript received 25 April 2023; revised 29 August 2023; accepted 10
October 2023. This work was supported in part by Guangzhou Municipal Science and Technology Project under Grant 2023A03J0011, in part by Guangdong Provincial Key Laboratory of Integrated Communications, Sensing and Computation for Ubiquitous Internet of Things, in part by National Foreign Expert Project, Project Number G2022030026L, in part by Science and Technology Development Fund of Macau SAR under Grant 0158/2022/A, and in part by the Guangdong Basic and Applied Basic Research Foundation (2022A1515011287). \textit{(Corresponding author: Wenhan Xu.)}}

\thanks{W. Xu and D.H.K. Tsang are with the Internet of Things Thrust, The Hong Kong University of Science and Technology (Guangzhou), Guangzhou, Guangdong 511400, China, and also with the Department of Electronic and Computer Engineering, The Hong Kong University of Science and Technology, Clear Water Bay, Hong Kong SAR, China (Email: wxube@ust.hk; eetsang@ust.hk).}
\thanks{J. Yu is with the Internet of Things Thrust, The Hong Kong University of Science and Technology (Guangzhou), Guangzhou, Guangdong 511400, China (Email: jiadongyu@hkust-gz.edu.cn).}
\thanks{Y. Wu is with The State Key Lab of Internet of Things for Smart City, and also with the Department of Computer and Information Science, The University of Macau, Macao SAR, China (Email: yuanwu@um.edu.mo).}
\thanks{Digital Object Identifier 10.1109/TGCN.}
}

%
%


\markboth{To appear in IEEE Transactions on Green Communications and Networking. Copyright was transferred TO IEEE. DOI: 10.1109/TGCN.2023.3330247.}%
{Shell \MakeLowercase{\textit{et al.}}: Bare Demo of IEEEtran.cls for IEEE Journals}

%



\maketitle

\begin{abstract}

The explosive development of the Internet of Things (IoT) has led to increased interest in mobile edge computing (MEC), which provides computational resources at network edges to accommodate computation-intensive and latency-sensitive applications. Intelligent reflecting surfaces (IRSs) have gained attention as a solution to overcome blockage problems during the offloading uplink transmission in MEC systems. This paper explores IRS-aided multi-cell networks that enable servers to serve neighboring cells and cooperate to handle resource exhaustion. We aim to minimize the joint energy and latency cost by jointly optimizing the computation tasks, edge computing resources, user beamforming, and IRS phase shifts.
The problem is decomposed into two subproblems—the MEC subproblem and the IRS communication subproblem—using the block coordinate descent (BCD) technique. The MEC subproblem is reformulated as a nonconvex quadratic constrained problem (QCP), while the IRS communication subproblem is transformed into a weight-sum-rate problem with auxiliary variables. We propose an efficient algorithm to alternately optimize the MEC resources and IRS communication variables.
Numerical results show that our algorithm outperforms benchmarks and that multi-cell MEC systems achieve additional performance gains when supported by IRS.

\end{abstract}


%
\IEEEpeerreviewmaketitle

\section{Introduction}

In recent years, the substantial growth of mobile devices (MDs) with limited memory space and computation power in the Internet-of-Things (IoT) era motivates the development of novel computational architectures \cite{mao2017survey,sun2016edgeiot}. As an emerging network structure, mobile edge computing (MEC) pushes abundant computational resources to the edges of the networks \cite{you2016energy}. This distributed computing paradigm tends to shorten the communication distance to meet the growing demand for novel intelligent applications such as augmented reality and virtual reality \cite{computing2014mobile,hu2015mobile,yang2022intelligent,mach2017mobile,wu2018noma}. 

Although MEC can effectively enable various latency-sensitive and computation-intensive services, the blockage of the line of sight (LoS) links between MDs and access points (APs) prevents the system from unleashing its full potential \cite{chu2020intelligent,yu2022deep}. As an emerging technology, intelligent reflecting surfaces (IRSs) can proactively reconfigure the wireless propagation channels to alleviate the blockage problems of LoS links via intelligent coordination of the reflecting signals\cite{shao2022target,shi2022intelligent}. IRS is comprised of a large number of low-cost passive reflecting elements, each of which can be adjusted in real-time to tune the amplitude and phase of the reflecting signals without using costly and power-hungry radio frequency (RF) chains\cite{yu2021irs,pan2022sum}. Since IRS can tackle the propagation-induced impairment and interference issues in the wireless channel by building the virtual LoS links, deploying IRS in MEC system dramatically improves the offloading performance of MDs when the communication environment is complex \cite{wu2021intelligent}. Specifically, MDs can offload their tasks to the MEC server without incurring higher energy consumption through high-capacity uplink wireless channels provided by IRS with reduced transmission latency.

A key challenge raised by IRS-aided MEC is balancing the limited energy and computing resources of the system with the stringent latency requirements of their tasks. Although the computation resource of the MEC server with reliable power is sufficient compared to MDs, the demand for edge computing remains unpredictable with the rapidly varying network environments. Thus, to tackle the resource exhaustion challenge, MEC servers in neighboring cells can be activated to serve users at the cell-edge in a collaborative manner. Moreover, the adjacent cells will reuse the same frequency resources, leading to severe inter-cell interference for the users at the cell-edge. With the development of 5G technologies and the widespread deployment of base stations, the density of base stations is increasing and is expected to reach up to 50 base stations per square kilometer \cite{ge20165g}. The high base station density creates a multicell MEC environment where users may simultaneously be within the overlapped coverage of multiple base stations \cite{chen2022dynamic}. However, to our best knowledge, it is still a challenging yet open issue on the multi-cell IRS-aided MEC, which usually leads to complicated resource allocation problems.

\subsection{Related Work}

Since the channel capacity for task offloading is critical to the performance of MEC, researchers introduced novel communication techniques to support MEC with low latency and high energy efficiency \cite{yu2022deep}. IRS is one of the promising communication techniques that can construct virtual LoS links when the LoS paths are blocked \cite{yang2022intelligent}. IRS uses massive reflecting elements to improve the offloading efficiency of MEC by enhancing both the latency and energy performance.
\cite{hua2019reconfigurable,cao2019intelligent} both considered applying IRS in MEC system to enhance the edge computing performance.
Specifically, an IRS-aided edge inference system was investigated in \cite{hua2019reconfigurable}, where the allocation of inference tasks, downlink transmit beamforming, and phase shift of the IRS were jointly optimized.
In \cite{cao2019intelligent}, a distributed optimization algorithm was proposed to solve the joint power control and passive beamforming optimization problem of MEC in IRS-mmWave systems.
In contrast to conventional IRS-aided MEC systems, \cite{huang2021reconfigurable} considered the complex offloading tasks, in which a system that simultaneously executed machine learning tasks both at the MEC server and the users was proposed to optimize the learning performance. Furthermore, \cite{huang2021reconfigurable2} extended the method in \cite{huang2021reconfigurable} into a more general model including heterogeneous learning tasks for broader application prospects.

Latency-efficiency \cite{bai2020latency,liu2021latency,zhou2020delay} and energy-efficiency \cite{li2021energy,huang2022integrated,xu2022energy,hua2021reconfigurable} are two main metrics that have been discussed in the IRS-aided MEC system. IRS has been proven beneficial in reducing the latency of MEC networks \cite{bai2020latency}. Furthermore, an IRS-aided device-to-device (D2D) offloading system was proposed to reduce the computation latency in the MEC system \cite{liu2021latency}. \cite{zhou2020delay} proposed a time-sharing method that allowed users to flexibly transmit their data via non-orthogonal multiple access (NOMA) or time division multiple access (TDMA) in an IRS-aided MEC system for delay optimization. Apart from the work that focused on latency optimization in the IRS-aided MEC system, a well-designed algorithm with the optimization objective of energy consumption was proposed in an IRS-aided single-cell multi-user MEC system through NOMA transmission \cite{li2021energy}.\cite{huang2022integrated} leveraged advanced IRS to improve both the energy performance of the radar sensing and MEC. An IRS-aided green edge inference system is considered in \cite{hua2021reconfigurable}, where the inference tasks generated from MDs are uploaded to BS and an overall power consumption minimization problem is formulated. 

By considering the cell-edge users in multi-cell networks that can offload their computing tasks to multiple MEC servers, the offloading efficiency can be further enhanced \cite{poularakis2019joint,liang2021multi,chen2022dynamic,liang2022two}. 
\cite{poularakis2019joint} proposed an algorithm with close-to-optimal performance using randomized rounding to jointly optimize the deployment of MEC servers and routing requirements in multi-cell networks. Furthermore, \cite{liang2021multi} developed an efficient relaxation-and-rounding-based solution for multi-cell MEC that can alleviate an overloaded MEC server by migrating its load to the nearby servers. \cite{chen2022dynamic} proposed a Lyapunov optimization-based online algorithm to solve the resource allocation problem in multi-cell networks by adaptively balancing the service migration cost and system performance.

In multi-cell networks, IRS can effectively improve the communication quality of cell-edge users when it is deployed at the cell boundary, as IRS can tackle severe co-channel interference issues from neighboring cells \cite{zhang2021intelligent}. \cite{cai2021intelligent} considered an IRS-assisted multi-cell multi-band system to minimize the total transmit power, in which different frequency bands are used by different BSs. Differently, \cite{pan2020multicell} considered the single frequency band that deployed an IRS at the cell boundary of multiple cells to assist the downlink transmission to the cell-edge users with reduced inter-cell interference and proposed a block coordinate descent (BCD) aided algorithm. \cite{zhang2021reconfigurable} proved that the IRS-aided multi-cell NOMA network showed superior performance than the system without IRS. Moreover, \cite{ni2021resource} optimized both the sum rate and the energy efficiency in multi-cell IRS-aided NOMA networks. \cite{hua2020intelligent} aimed at maximizing the minimum achievable rate in multi-cell system by jointly optimizing the precoding matrix at the BSs and the phase shifts at the IRS while taking into account the fairness among cell-edge users.

\subsection{Motivation and Contribution}

Although IRS-aided MEC has been studied \cite{bai2021resource,hua2019reconfigurable,cao2019intelligent,bai2020latency,liu2021latency,zhou2020delay,li2021energy,yu2022irs,xu2022deep}, there are few works focusing on activate MEC servers in other neighboring cells in IRS-aided MEC system. 
In conventional MEC systems, MDs can only offload their computing tasks to one MEC server. It is challenging to activate multiple MEC servers to collaboratively serve a single user to tackle resource exhaustion because of the complicated joint resource allocation problems. Moreover, cell-edge deployment of IRS can alleviate severe co-channel interference from neighboring cells \cite{rezaei2022energy}, which motivates us to explore the multi-cell system with multiple MEC servers.
Although \cite{chen2022dynamic,poularakis2019joint,liang2021multi} explored the multi-cell MEC for service migration, they did not consider the LoS links blockage problem that can prevent the MEC systems from unleashing their full potential. 
However, the existing IRS-aided MEC works mainly considered either execution latency \cite{bai2020latency,liu2021latency,zhou2020delay} or energy consumption \cite{li2021energy} in MEC systems, lacking joint optimization of both critical objectives.

Motivated by the aforementioned literature review, we consider the IRS-aided MEC in multi-cell networks, and both energy consumption and computing latency are designed as the optimization objectives. The main contributions are summarized as follows.

\begin{enumerate}
    \item We develop an IRS-aided MEC model in multi-cell networks, which enables the cell-edge users to offload their computing tasks to several MEC servers at different BSs. We design the weighted optimization metric that consists of both computing latency and energy consumption. We formulate the minimization problem in the IRS-aided multi-cell MEC system by optimizing the computation offloading volume, the edge computing resources allocated to each device, the beamforming vector, and the phase-shifting matrix.
    \item The problem is decomposed into a MEC subproblem and an IRS communication subproblem through the BCD technique. Specifically, the MEC subproblem is transformed into a standard quadratic constrained problem (QCP). The IRS communication subproblem is transformed into a weight-sum-rate problem with the assistance of auxiliary variables.
    \item We propose the algorithm named as BCD-FP-DC. Decomposed by BCD, the MEC subproblem in the nonconvex QCP form is solved by a spatial branch-and-bound method. The IRS communication subproblem in the weight-sum-rate form is solved by the fractional programming (FP) technique, where the difference-of-convex (DC) problem is then solved by majorization minimization (MM) method.

    \item We present the numerical results to validate the performance of our proposed BDC-FD-DC algorithm. The results show that our BCD-FP-DC algorithm can achieve better performance with a lower system cost than all the benchmarks under the large size of the IRS elements. Moreover, the multi-cell IRS-aided MEC framework can achieve additional performance gains compared to the multi-cell MEC system without the support of the IRS.
    
\end{enumerate}

\subsection{Organization and Notation}

This paper is organized as follows. In Section II, the IRS-aided MEC system model in multi-cell is introduced, and the cost minimization problem is formulated. In Section III, the proposed BCD-FP-DC algorithm is presented. In Section IV and Section V, the numerical results and conclusions are presented, respectively.

Bold lowercase and uppercase letters denote vectors and matrices, respectively. $\Vert \boldsymbol{x}\Vert$ refers to the 2-norm of vector $\boldsymbol{x}$. $\nabla f(\boldsymbol{x})$ returns the gradient of the function $f$. $diag(\boldsymbol{x})$ returns a diagonal matrix with the elements of vector $\boldsymbol{x}$ on the main diagonal. $|\mathbf X|$, $\mathbf X^T$, $\mathbf X^H$, and $\mathrm{Tr}[\mathbf X]$ refer to the determinant, transpose, conjugate transpose, and trace of a matrix $\mathbf X$, respectively.

\begin{table}[t]
\renewcommand{\arraystretch}{1.3}
\caption{Parameters notation}
\label{Parameters_notation}
\centering
\begin{tabular}{c|c}
\hline
\bfseries Parameters & \bfseries Notation\\
\hline
$q,k$ & $q^{th}$ BS and $k^{th}$ user\\
$\mathbf{s}_{q,k}$ & Symbol vector transmitted to BS\\
$\mathbf{F}_{q,k}$ & Beamforming vector at the user\\
${\mathbf H}_{q,k}$ & Baseband channel from user to BS\\
${\mathbf G}_{q,R}$ & Baseband channel from IRS to BS\\
${\mathbf H}_{R,k}$ & Baseband channel from user to IRS\\
$\mathbf{\Phi}$ & Diagonal phase-shifting matrix of IRS\\
$\theta$ & Phase shift of IRS elements\\
$\mathbf{n}_{q}$ & Noise vector\\
$\sigma^2$ & Variance of noise\\
$R_{q,k}$ & Achievable data rate (nat/s/Hz)\\
${\mathbf J}$ & Interference-plus-noise covariance matrix\\
$B$ & channel bandwidth\\
$L_k$ & Total number of bits to be processed\\
$\ell_{q,k}$ & Number of bits to be offloaded\\
$c_k$ & Number of CPU cycles required to process a single bit\\
$D_k$ & Time required for computation\\
$E_k$ & Energy consumption\\
$f^{\mathbb L}_k$ & Computational capability at the user\\
$f^{\mathbb E}_{q,k}$ & Computational capability allocated by MEC server\\
$P_{q,k}^{\mathbb E}$ & Transmit power\\
$C_{k}$ & System cost\\
$\zeta$ & Weight factor between energy consumption and latency\\
$\omega_{k}$ & Weight of the user\\
\hline
\end{tabular}
\end{table}

\section{System Model}

\begin{figure}[t]
\centering
\includegraphics[width=0.47\textwidth]{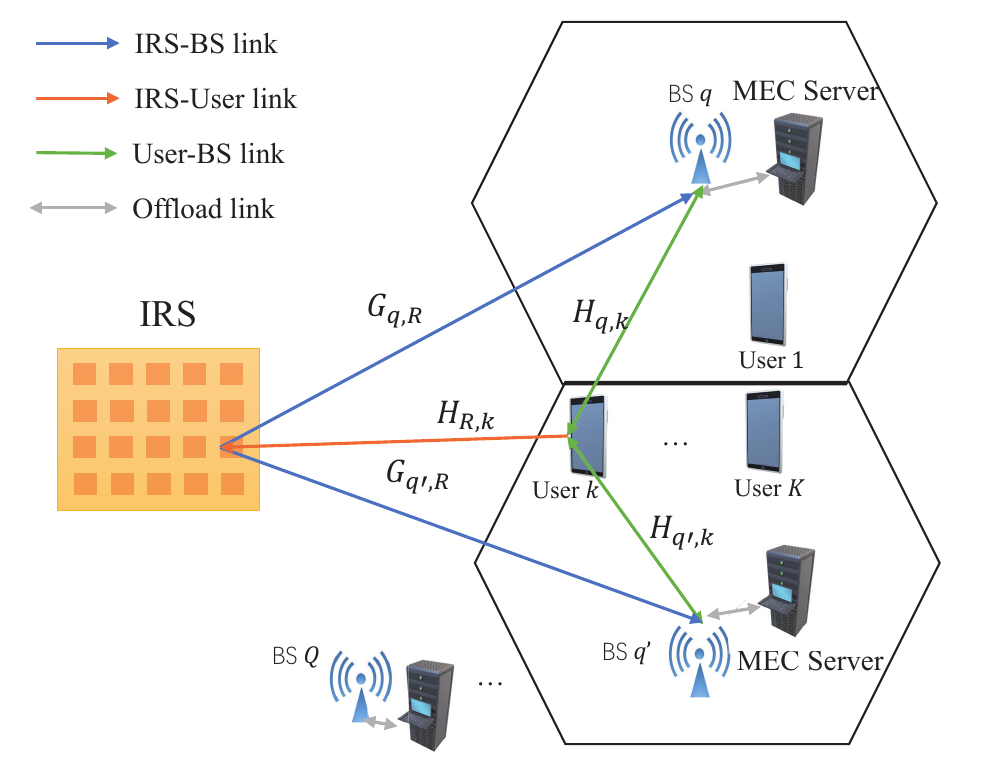}
\caption{System model for a multi-cell IRS-aided MEC system with $K$ cell-edge users and $Q$ cells}
\label{IRS_MEC_Multicell_model}
\end{figure}

As illustrated in Fig. \ref{IRS_MEC_Multicell_model}, we consider the uplink IRS-aided MEC in a multi-cell networks system. In each macro cell, there is a single BS connected with an MEC server that serves $K$ cell users. Table \ref{Parameters_notation} summarizes the important symbols used in this paper.

\subsection{Communication Model}
As shown in Fig. \ref{IRS_MEC_Multicell_model}, there are $K$ cell-edge users and $Q$ cells. Each BS and each user in our IRS-aided MEC in multi-cell networks system have $N_{BS}\geq 1$ and $N_{U}\geq 1$ antennas, respectively. We employ an IRS with $M$ reflection elements at the cell-edge to enhance the spectral and energy efficiency across the whole system by carefully designing the reflecting phase shift.

For the uplink communication, $\mathbf{s}_{q,k}$ is the signal transmitted to the $q^{th}$ BS from the $k^{th}$ cell-edge user. The signal transmitted by the $k^{th}$ cell-edge user is given by
\begin{equation} 
	{\mathbf x}_k = \sum \limits _{q = 1}^{Q} {{\mathbf{F}}_{q,k}{\mathbf{s}}_{q,k}},
	\label{signal_transmitted_by_the_lth_BS}
\end{equation}
where $\mathbf{F}_{q,k}$ is the beamforming vector used by the $k^{th}$ cell-edge user for transmitting the data vector $\mathbf{s}_{q,k}$ to the $q^{th}$ BS. 

With the assumption that the channel state information (CSI) is perfectly known by the system controller \cite{wu2019towards,zheng2019intelligent}, we focus on the joint optimization of the latency and energy consumption of the system. The received signal vector at the $q^{th}$ BS can be written as
\begin{equation}
	{{\mathbf y}_q} = \underbrace {\sum \limits _{k = 1}^K {{{\mathbf H}_{q,k}}{{\mathbf x}_k}} }_{{\mathrm{From~Users}}} + \underbrace {\sum \limits _{k = 1}^K { {\mathbf G}_{q,R} \boldsymbol {\Phi } {\mathbf H}_{R,k} } {{\mathbf x}_k}}_{{\mathrm{From~IRS}}} +{{\mathbf n}_q}
	\label{received_signal_vector_1},
\end{equation}
where $\mathbf{\Phi}=diag\{ e^{j\theta_1},e^{j\theta_2},...,e^{j\theta_M}\}$ is the diagonal phase-shifting matrix of the IRS, $\mathbf{n}_{q}$ is the noise vector with variance as $\sigma^2$, and $\theta_n$ denotes the phase shift of the $n^{th}$ reflecting element on the IRS \cite{lu2021aerial}. As shown in Fig. \ref{IRS_MEC_Multicell_model}, the CSI for user-BS, IRS-BS, and IRS-user are denoted by ${\mathbf H}_{q,k}$, ${\mathbf G}_{q,R}$, and ${\mathbf H}_{R,k}$, respectively. We introduce $\bar{\mathbf{H}}_{q,k}$ to simplify the expressions, which is defined as
\begin{equation}
	\bar{\mathbf{H}}_{q,k}\triangleq\mathbf{H}_{q,k} + {\mathbf G}_{q,R} \boldsymbol {\Phi } {\mathbf H}_{R,k}.
\end{equation}
Hence, $\mathbf{y}_{q}$ can be separated into the signal part and the interference part, which can be written as
\begin{equation}
\begin{aligned}
	\mathbf{y}_{q} = 
	& \sum \limits _{k = 1}^K \sum \limits _{n = 1}^Q{\bar {\mathbf H}_{q,k}{\mathbf F}_{n,k}{\mathbf s}_{n,k}       }  +{\mathbf n}_{q}\\
	=& \underbrace {\sum \limits _{k = 1}^K {\bar {\mathbf H}_{q,k}{\mathbf F}_{q,k}{\mathbf s}_{q,k}       } }_{{\mathrm {Signal}}} +    \underbrace {\sum \limits _{k = 1}^K \sum \limits _{n = 1,n \ne q}^Q{\bar {\mathbf H}_{q,k}{\mathbf F}_{n,k}{\mathbf s}_{n,k}       } }_{{\mathrm {Interference}}}   +{\mathbf n}_{q}.
\end{aligned}
\end{equation}
Therefore, the achievable data rate of the $k^{th}$ cell-edge user from the $q^{th}$ BS can be written as
\begin{equation}
{R_{q,k}}= {\log }\left |{ {{\mathbf{I}} + {{\bar {\mathbf {H}}}_{q,k}}{{\mathbf{F}}_{q,k}}{\mathbf{F}}_{q,k}^{\mathrm{H}}\bar {\mathbf {H}}_{q,k}^{\mathrm{H}}{\mathbf{J}}_{q,k}^{ - 1}} }\right |,
\end{equation}
where the interference-plus-noise covariance matrix $\mathbf{J}_{q,k}$ can be formulated as
\begin{equation}
\begin{aligned}
	{{\mathbf{J}}_{q,k}} = 
	& \sum \limits _{m = 1,m \ne k}^{K} {{{\bar {\mathbf {H}}}_{q,m}}{{\mathbf{F}}_{q,m}}{\mathbf{F}}_{q,m}^{\mathrm{H}}\bar {\mathbf {H}}_{q,m}^{\mathrm{H}}} \\
	&{+ \sum \limits _{m = 1}^{K} {\sum \limits _{n = 1,n \ne q}^{Q} {{{\bar {\mathbf {H}}}_{q,k}}{{\mathbf{F}}_{n,m}}{\mathbf{F}}_{n,m}^{\mathrm{H}}\bar {\mathbf {H}}_{q,k}^{\mathrm{H}}} } + {\sigma ^{2}}{\mathbf{I}}}.
\end{aligned}
\end{equation}
Therefore, the achievable data rate of the $k^{th}$ cell-edge user from the $q^{th}$ BS can be calculated with the given beamforming vector $\mathbf{F}_{q,k}$, and the diagonal phase-shifting matrix $\mathbf{\Phi}$ of the IRS.

\subsection{Computing Model}

We consider that the computing tasks in this paper are data-partitioning-based applications. In these data partition-oriented application tasks, a fraction of them can be offloaded to the MEC server, and the rest can be processed locally, which leads to local computing and edge computing analyzed as the following.

\subsubsection{Local Computing}
In terms of data partition-oriented application, the latency imposed by local computation at the $k^{th}$ cell-edge user is
\begin{equation}
	D_k^{\mathbb L}=(L_{k}-\sum ^{Q}_{q=1}\ell _{q,k})\frac{c_k}{f^{\mathbb L}_k},
\end{equation}
where $L_k$ is the total number of bits to be processed, $\ell_{q,k}$ is the number of bits offloaded to the edge computing server at the $q^{th}$ BS, $c_k$ is the number of CPU cycles required to process a single bit, and $f^{\mathbb L}_k$ is the computational capability (CPU cycles per second) at the users.

The energy consumption per CPU cycle \cite{liu2020resource} at the $k^{th}$ cell-edge user is denoted as $E_k^d$. Then, the total local energy consumption of the task computation at the $k^{th}$ cell-edge user is denoted as
\begin{equation}
    E_k^{\mathbb L}=c_kE_k^d(L_k-\sum ^{Q}_{q=1}\ell _{q,k}).
	\label{local_consumption}
\end{equation}
\subsubsection{Edge Computing}
The total latency $D_{q,k}^{\mathbb E}$ consists of the computation offloading latency, the edge computing latency, and the result transmitting latency. Usually, the computation result is simple and the result transmitting latency can be ignored upon using the technique of ultra-reliable low-latency communications \cite{mao2017survey,bai2020latency}. Therefore, $D_{q,k}^{\mathbb E}$ is given by
\begin{equation}
	D_{q,k}^{\mathbb E}=\frac{\ell _{q,k}}{BR_{q,k}}+\frac{\ell _{q,k} c_k}{f^{\mathbb E}_{q,k}},
\end{equation}
where $B$ is the bandwidth of the channel, $\ell_{q,k}$ is the number of bits offloaded to the edge computing server at the $q^{th}$ BS, $f_{q,k}^{\mathbb E}$ is the computational capability (CPU cycles per second) allocated to the $k^{th}$ device by the edge computing server at the $q^{th}$ BS. Each edge server has a computing capacity constraint $\sum ^{K}_{k=1} f^{\mathbb E}_{q,k} \leq f^{\mathbb E}_{q,\text{total}}$. The overall latency $D_k^{\mathbb E}$ of the $k^{th}$ device for edge computing is the maximum among different MEC servers, which is expressed as
\begin{equation}
	D_k^{\mathbb E}=\max \big \{ D_{1,k}^{\mathbb E}, D_{2,k}^{\mathbb E},...,D_{Q,k}^{\mathbb E}\big \}.
\end{equation}

Then, the total edge energy consumption $E_k^{\mathbb E}$ of a single computation task for the cell-edge user $k$ is the sum of the computing energy consumption and the transmitting energy consumption \cite{liu2020resource}, which is expressed as
\begin{equation}
    E_k^{\mathbb E}=\sum ^{Q}_{q=1}{c_kE_{q}^s\ell_{q,k}}+\sum ^{Q}_{q=1}{P^{\mathbb E}_{q,k}\frac{\ell_{q,k}}{BR_{q,k}}}.
	\label{edge_consumption}
\end{equation}
In eq. (\ref{edge_consumption}), $P_{q,k}^{\mathbb E}$ denotes the transmit power from $k^{th}$ cell-edge user to $q^{th}$ BS, and $E_{q}^s$ denotes the energy consumption per CPU cycle at the $q^{th}$ BS.
\subsection{Problem Formulation}

The latency of the $k^{th}$ cell-edge user can be readily calculated by selecting the maximum value of both local and edge computing time, which is given by
\begin{equation}
	\begin{aligned}
		D_k&=\max \big \{ D^{\mathbb L}_{k}, D^{\mathbb E}_{k} \big \} \\
  &=\max \left\{ (L_{k}-\sum ^{Q}_{q=1}\ell _{q,k})\frac{c_k}{f_k^{\mathbb L}}, \max_{\forall q}\{\frac{\ell _{q,k}}{BR_{q,k}}+\frac{\ell _{q,k} c_k}{f^{\mathbb E}_{q,k}}\}\right\}.
	\end{aligned}
    \label{D_k}
\end{equation}
The total energy consumption for the $k^{th}$ cell-edge user is the sum of local consumption in eq. (\ref{local_consumption}) and edge consumption in eq. (\ref{edge_consumption}), which is given by
\begin{equation}
    E_k=c_kE_k^d(L_k-\sum ^{Q}_{q=1}\ell _{q,k})+\sum ^{Q}_{q=1}{c_kE_{q}^s\ell_{q,k}}+\sum ^{Q}_{q=1}{\frac{P^{\mathbb E}_{q,k}\ell_{q,k}}{BR_{q,k}}}.
    \label{E_k}
\end{equation}
The energy consumption and the task execution latency are two main costs in the edge computing network. We introduce the weight factor between the energy consumption and the task execution latency, which combines different types of functions with different units into a weighted cost function. Therefore, the cost function $C_{k}$ for the $k^{th}$ cell-edge user can be defined as
\begin{equation}
    \begin{aligned}
		C_{k}&=E_k+\zeta D_k\\
		&=c_kE_k^d(L_k-\sum ^{Q}_{q=1}\ell _{q,k})+\sum ^{Q}_{q=1}{c_kE_{q}^s\ell_{q,k}}+\sum ^{Q}_{q=1}{\frac{P^{\mathbb E}_{q,k}\ell_{q,k}}{BR_{q,k}}}\\
		&+\zeta\max \left\{ (L_{k}-\sum ^{Q}_{q=1}\ell _{q,k})\frac{c_k}{f_k^{\mathbb L}}, \max_{\forall q}\{\frac{\ell _{q,k}}{BR_{q,k}}+\frac{\ell _{q,k} c_k}{f^{\mathbb E}_{q,k}}\}\right\},
	\end{aligned}
	\label{cost_function}
\end{equation}
where $\zeta$ is the weight factor between the energy consumption and the latency. The physical meaning of $\zeta$ is to provide a mechanism to balance these two different dimensions within the system, considering that energy consumption and execution latency may have different units and significance in various contexts. By adjusting the value of $\zeta$, one can tailor the system's optimization towards either minimizing energy consumption or reducing execution time, depending on the specific objectives and constraints of the application. For instance, a higher value of $\zeta$ indicates that the system places more emphasis on reducing latency, while a lower value suggests a preference for energy efficiency. This weighted approach allows for a unified framework to evaluate and optimize the system performance concerning both energy and time. In our problem, $\zeta$ is assumed to be known in advance.

Our objective is to minimize the weighted cost function of all the cell-edge users by jointly optimizing the computation offloading volume $\boldsymbol\ell$, the edge computing resources $\boldsymbol f^{\mathbb E}$ allocated to each device, the beamforming vector $\mathbf{F}$, and the phase-shifting $\boldsymbol{\theta}$ of the IRS. Therefore, we can formulate the system-cost-minimization Problem $\mathcal P1$:
\begin{subequations}
    \label{P0-1}
    \begin{align}
		\mathcal P&1:\min \limits _{\mathbf{F},\boldsymbol{\theta},\boldsymbol \ell,\boldsymbol f^{\mathbb E}} \sum ^{K}_{k=1} \omega_{k} C_{k} \nonumber\\
		\text{s.t.}\ &\Vert\mathbf{F}_{q,k}\Vert < 1, \forall q,k,\label{P0-1_8}\\
		&0 \leq \theta _{n} < 2\pi, \forall n,\label{P0-1_5}\\
		&0\leq \sum ^{Q}_{q=1}\ell _{q,k}\leq L_{k}, \forall k,\label{P0-1_6}\\
		&0\leq \sum ^{K}_{k=1} f^{\mathbb E}_{q,k} \leq f^{\mathbb E}_{q,\text{total}},\forall q,\label{P0-1_7}
	\end{align}
\end{subequations}
where $\omega_{k}$ represents the weight of the $k^{th}$ cell-edge user originating from the system settings. Constraint (\ref{P0-1_8}) specifies the range of the beamforming vector. Constraint (\ref{P0-1_5}) specifies the range of the phase shift. Constraints (\ref{P0-1_6}) and (\ref{P0-1_7}) restrict the computation offloading variables and the edge computing resources allocated to each user, respectively.

\section{Proposed Algorithm}

In this section, we use the BCD technique to tackle this nonconvex problem. BCD method is applied to alternatively optimize the IRS communication decision variables ($\mathbf{F}$ and $\boldsymbol{\theta}$) and the MEC decision variables ($\boldsymbol f^{\mathbb E}$ and $\boldsymbol\ell$) in the original Problem $\mathcal P1$. The BCD approach is described as Algorithm \ref{BCD_algorithm}.

\begin{algorithm}[t]
	\caption{BCD structure}
	\begin{algorithmic}[1]
		\ENSURE ${\mathbf H}_{q,k},{\mathbf G}_{q,R},{\mathbf H}_{R,k}, L_{k}, f^{\mathbb E}_{q,\text{total}}, \omega_{k}, \zeta, B, \sigma^2$
		\REQUIRE $\mathbf{F},\boldsymbol{\theta},\boldsymbol\ell,\boldsymbol f^{\mathbb E}$
		\STATE Set $n=1$, Calculate ${Cost}^n=\sum ^{K}_{k=1} \omega_{k} C_{k}$.
		\STATE Initialize all optimization variables $\mathbf{F},\boldsymbol{\theta},\boldsymbol\ell,\boldsymbol f^{\mathbb E}$ with random values.
		\WHILE {${Cost}^n-{Cost}^{n-1}>\epsilon$, \AND $n<N$}
		    \STATE Update $n=n+1$.
			\STATE Fix $\mathbf{F}$ and $\boldsymbol{\theta}$, and update $\boldsymbol\ell$ and $\boldsymbol f^{\mathbb E}$ by solving the MEC subproblem.
			\STATE Fix $\boldsymbol\ell$ and $\boldsymbol f^{\mathbb E}$, and update $\mathbf{F}$ and $\boldsymbol{\theta}$ by solving the IRS communication subproblem.
			\STATE Calculate ${Cost}^n=\sum ^{K}_{k=1} \omega_{k} (E_k+\zeta D_k)$ by eq. (\ref{cost_function}).
		\ENDWHILE
	\end{algorithmic}
	\label{BCD_algorithm}
\end{algorithm}

\subsection{MEC Subproblem}

The MEC subproblem of the BCD structure is described as Problem $\mathcal P2$ while fixing the IRS communication setting:
\begin{equation}
    \label{BCD_FP_DCP_2}
    \begin{aligned}
		\mathcal P2:&\min \limits _{\boldsymbol{f^{\mathbb E},\ell}} \sum ^{K}_{k=1} \omega_{k} C_{k}\\
		\text{s.t.}\ &\text{(\ref{P0-1_6})(\ref{P0-1_7})},
	\end{aligned}
\end{equation}
To solve the Problem $\mathcal P2$, we simplify the original Problem $\mathcal P2$ by relaxing $E_k$ and $D_k$ to remove the maximum function and the optimization variables will be expanded to $\{\boldsymbol\ell,\boldsymbol E,\boldsymbol D,\boldsymbol f^{\mathbb E}\}$, where $\boldsymbol E = \{E_k, \forall k\}$ and $\boldsymbol D = \{D_k, \forall k\}$. Eq. (\ref{E_k}) is relaxed as
\begin{equation}
    c_kE_k^d(L_k\!-\!\sum ^{Q}_{q=1}\ell _{q,k})+\sum ^{Q}_{q=1}{c_kE_{q}^s\ell_{q,k}}+\sum ^{Q}_{q=1}{P^{\mathbb E}_{q,k}\frac{\ell_{q,k}}{BR_{q,k}}}\!\leq\!E_k, \forall k\label{BCD_FP_DCP_1_1}.
\end{equation}
Moreover, eq. (\ref{D_k}) is relaxed as
\begin{equation}
    (L_{k}-\sum ^{Q}_{q=1}\ell _{q,k}) \frac{c_{k}}{f^{\mathbb L}_{k}}\leq D_k, \forall k\label{BCD_FP_DCP_1_2},
\end{equation}
and
\begin{equation}
    \frac{\ell _{q,k}}{BR_{q,k}}+\frac{\ell _{q,k} c_k}{f^{\mathbb E}_{q,k}}\leq D_k, \forall q,k\label{BCD_FP_DCP_1_3}.
\end{equation}
Therefore, the simplified problem without requiring any maximum functions is formulated as
\begin{equation}
    \label{BCD_FP_DCP_1}
    \begin{aligned}
		\mathcal P3:&\min \limits _{\boldsymbol{\ell,E,D,f^{\mathbb E}}} \sum ^{K}_{k=1} \omega_{k} (E_k+\zeta D_k)\\
		\text{s.t.}\ &\text{(\ref{BCD_FP_DCP_1_1})(\ref{BCD_FP_DCP_1_2})(\ref{BCD_FP_DCP_1_3})(\ref{P0-1_6})(\ref{P0-1_7})}.
	\end{aligned}
\end{equation}

To solve the Problem $\mathcal P3$, as all the constraints are either quadratic or linear, we can transform it into a standard QCP. Then the equivalent problem can be formulated as
\begin{subequations}
    \label{BCD_FP_DCP_3}
	\begin{align}
		\mathcal P3a:&\min \limits _{\boldsymbol{\ell,E,D,f^{\mathbb E}}} \sum ^{K}_{k=1} \omega_{k} (E_k+\zeta D_k) \nonumber\\
		\text{s.t.}\ &\sum ^{Q}_{q=1}{c_kE_{q}^s\ell_{q,k}}-c_kE_k^d\sum ^{Q}_{q=1}\ell _{q,k}\nonumber\\
        &+\sum ^{Q}_{q=1}{\frac{P^{\mathbb E}_{q,k}}{BR_{q,k}}\ell_{q,k}}-E_k+c_kE_k^dL_k\leq 0, \forall k, \label{BCD_FP_DCP_3_1}\\
		&\frac{1}{BR_{q,k}}\ell _{q,k}f^{\mathbb E}_{q,k}-D_kf^{\mathbb E}_{q,k}+\ell _{q,k}c_k\leq 0, \forall q,k, \label{BCD_FP_DCP_3_2}\\
		&-D_k-\frac{c_{k}}{f^{\mathbb L}_{k}}\sum ^{Q}_{q=1}\ell _{q,k}+\frac{c_{k}L_k}{f^{\mathbb L}_{k}}\leq 0, \forall k, \label{BCD_FP_DCP_3_3}\\
		&\text{(\ref{P0-1_6})(\ref{P0-1_7})}.\nonumber
	\end{align}
\end{subequations}
To solve the non-convex quadratic minimization problem, the QCP variable $\boldsymbol{x}$ is set as $\{\boldsymbol\ell,\boldsymbol E,\boldsymbol D,\boldsymbol f^{\mathbb E}\}$, and therefore, Problem $\mathcal P3a$ can be rewritten as a standard QCP form: 

\begin{subequations}
    \label{BCD_FP_DCP_4}
	\begin{align}
		\mathcal P3b:&\min \limits _{\boldsymbol{x}} c^T\boldsymbol{x}+d \nonumber\\
		\text{s.t.}\ &\frac{1}{2} \boldsymbol{x}^TQ_i\boldsymbol{x}+r_i^T\boldsymbol{x}+s_i\leq 0, i=1,2,...,m,\label{BCD_FP_DCP_4_1}\\
		&\text{(\ref{BCD_FP_DCP_3_1})(\ref{BCD_FP_DCP_3_3})(\ref{P0-1_6})(\ref{P0-1_7})},\nonumber
	\end{align}
\end{subequations}
where $c^T$, $d$, $Q_i$, $r_i^T$, and $s_i$ are constant vectors, and $Q_i$ are constant matrices. $\boldsymbol{x}^TQ_i\boldsymbol{x}$ is the symmetric bilinear form of the coefficient of the quadratic term in constraint (\ref{BCD_FP_DCP_3_2}). Constraint (\ref{BCD_FP_DCP_4_1}) is equivalent to constraint (\ref{BCD_FP_DCP_3_2}), and the others are linear. The reformulation from $\mathcal P3a$ to $\mathcal P3b$ and the specific values of $d$, $i$, $m$, $s_i$, $c^T$, $Q_i$ and $r_i^T$ are shown in Appendix \ref{P3a_P3b}\footnotemark{}.\footnotetext{Problem $\mathcal P3b$ is a nonconvex optimization problem because $Q_i$ in the quadratic constraints (\ref{BCD_FP_DCP_4_1}) are not positive semi-definite matrices. We use Gurobi \cite{gurobi2018gurobi} to solve subproblem $\mathcal P3b$ by translating them into a bilinear form for which a convex relaxation can be constructed. The reformulated problem is then solved using a spatial branch-and-bound algorithm \cite{smith1999symbolic}.}

\subsection{IRS Communication Subproblem}

While fixing the MEC setting, the IRS communication subproblem is described as
\begin{equation}
    \label{BCD_FP_DCP_5}
    \begin{aligned}
		\mathcal P4:&\min \limits _{\boldsymbol{\boldsymbol{\theta},\mathbf{F}}} \sum ^{K}_{k=1} \omega_{k} C_{k}\\
		\text{s.t.}\ &\text{(\ref{P0-1_8})(\ref{P0-1_5})}.
	\end{aligned}
\end{equation}
We introduce the auxiliary variable $\lambda_{q,k}=\frac{1}{R_{q,k}}$ in this problem. We notice that $\sum ^{K}_{k=1} \omega_{k} C_{k}=f(\lambda)$. We perform a Taylor expansion on the objective:
\begin{equation}
	\sum ^{K}_{k=1} \omega_{k} C_{k}=f(\lambda^0)+\sum^K_{k=1}\sum^Q_{q=1}\left(\frac{\partial f(\lambda)}{\partial \lambda_{q,k}}(\lambda_{q,k}-\lambda^0_{q,k})\right)+o(\lambda),
\end{equation}
where $o(\lambda)$ represents an infinitesimal term of higher order than $\lambda$. We introduce the auxiliary variable $\omega^{(1)}_{q,k}=\frac{\partial f(\lambda) }{\partial \lambda_{q,k}}$. After ignoring the higher-order small quantities and omitting the fixed term, the objective is transformed to
\begin{equation}
    \min\limits_{\boldsymbol{\boldsymbol{\theta},\mathbf{F}}} \sum^K_{k=1}\sum^Q_{q=1}\frac{\omega^{(1)}_{q,k}}{R_{q,k}}.
\end{equation}
We introduce the auxiliary variable $\beta_{q,k}=\frac{\omega^{(1)}_{q,k}}{R_{q,k}}$. Then the problem can be transformed as
\begin{subequations}
    \label{BCD_FP_DCP_5_1}
    \begin{align}
        \mathcal P4a:&\max \limits _{\boldsymbol{\boldsymbol{\theta},\mathbf{F}}} \sum^K_{k=1}\sum^Q_{q=1} \lambda_{q,k}\beta_{q,k}R_{q,k} \nonumber\\
		\text{s.t.}\ & 0 \leq \theta _{n} < 2\pi, \forall n,\\
		& \Vert\mathbf{F}_{q,k}\Vert < 1, \forall q,k.
	\end{align}
\end{subequations}
The first step is to obtain $\boldsymbol{\theta}$ and $\mathbf{F}$ by solving Problem $\mathcal P4a$. The second step is to update $\beta$ and $\lambda$ by using the modified Newton’s method in \cite{bai2020latency} until convergence. We further introduce the auxiliary weight $\omega_{q,k}^*$ updated by $\omega_{q,k}^*=\lambda_{q,k}\beta_{q,k}$. Hence, the weight-sum-rate maximization problem is defined as

\begin{subequations}
    \label{BCD_FP_DCP_6}
    \begin{align}
		\mathcal P4b: &\max \limits _{\boldsymbol{\boldsymbol{\theta},\mathbf{F}}} \sum^K_{k=1}\sum^Q_{q=1} \omega^*_{q,k} R_{q,k} \nonumber\\
		\text{s.t.}\ & R_{q,k}= \log\left |{ {{\mathbf{I}} + {{\bar {\mathbf {H}}}_{q,k}}{{\mathbf{F}}_{q,k}}{\mathbf{F}}_{q,k}^{\mathrm{H}}\bar {\mathbf {H}}_{q,k}^{\mathrm{H}}{\mathbf{J}}_{q,k}^{ - 1}} }\right |,\label{BCD_FP_DCP_6_1}\\
		& {\bar{\mathbf H}}_{q,k}=G_{q,R} \boldsymbol{\Phi} H_{R,k}^H +\mathbf{H}_{q,k},\label{BCD_FP_DCP_6_2}\\
		& \boldsymbol{\Phi}=diag\{ e^{j\theta_1},e^{j\theta_2},...,e^{j\theta_M}\},\label{BCD_FP_DCP_6_3}\\
		& 0 \leq \theta _{n} < 2\pi, \forall n.\label{BCD_FP_DCP_6_4}
	\end{align}
\end{subequations}
Constraint (\ref{BCD_FP_DCP_6_1}) can be rewritten as
\begin{equation}
	R_{q,k}=\log \left(1+\frac{\left\Vert{\bar{\mathbf H}}_{q,k}\mathbf{F}_{q,k}\right\Vert^2}{\sum^{Q,K}_{i=1,j=1,i\neq q}\left\Vert{\bar{\mathbf H}}_{i,k}\mathbf{F}_{i,j}\right\Vert^2+\sigma^2}\right).
\end{equation}
By adding the auxiliary vector $\boldsymbol\alpha$ introduced by the Lagrangian dual transform based on the equation (41) in \cite{shen2018fractional}, Problem $\mathcal P4b$ can be equivalently transformed as
\begin{subequations}
    \label{BCD_FP_DCP_7}
    \begin{align}
		\mathcal P4c: &\max \limits _{\boldsymbol{\boldsymbol{\theta},\mathbf{F},\alpha}} J=\frac{1}{\ln 2}\sum^K_{k=1}\sum^Q_{q=1} \omega^*_{q,k} \ln(1+\alpha_{q,k})-\omega^*_{q,k}\alpha_{q,k}\nonumber\\
        &+\frac{\omega^*_{q,k}(1+\alpha_{q,k})\gamma_{q,k}}{1+\gamma_{q,k}} \nonumber\\
		\text{s.t.}\ & R_{q,k}\leq{\log }(1+\gamma_{q,k}),\label{BCD_FP_DCP_7_1}\\
		& \gamma_{q,k}=\frac{\left\Vert{\bar{\mathbf H}}_{q,k}\mathbf{F}_{q,k}\right\Vert^2}{\sum^{Q,K}_{i=1,j=1,i\neq q}\left\Vert{\bar{\mathbf H}}_{i,k}\mathbf{F}_{i,j}\right\Vert^2+\sigma^2},\label{BCD_FP_DCP_7_2}\\
		&\text{(\ref{BCD_FP_DCP_6_2})(\ref{BCD_FP_DCP_6_3})(\ref{BCD_FP_DCP_6_4})}.\nonumber
	\end{align}
\end{subequations}
We get ${\alpha_{q,k}}=\gamma_{q,k}$ by setting $\frac{\partial J}{\partial \alpha_{q,k}}=0$. When given $\boldsymbol{\theta}$ and $\mathbf{F}$, $\alpha_{q,k}$ can be updated by (\ref{BCD_FP_DCP_7_1}) in each iteration. Given $\boldsymbol\alpha$, Problem $\mathcal P4c$ can be recast as
\begin{equation}
    \label{BCD_FP_DCP_8}
    \begin{aligned}
		\mathcal P4d: &\max \limits _{\boldsymbol{\boldsymbol{\theta},\mathbf{F}}} \sum^K_{k=1}\sum^Q_{q=1} \frac{\omega^*_{q,k}(1+\alpha_{q,k})\gamma_{q,k}}{1+\gamma_{q,k}}+\frac{\omega^*_{q,k}(1+\alpha_{q,k})\gamma_{q,k}}{1+\gamma_{q,k}} \\
		\text{s.t.}\ & \text{(\ref{BCD_FP_DCP_7_1})(\ref{BCD_FP_DCP_7_2})(\ref{BCD_FP_DCP_6_2})(\ref{BCD_FP_DCP_6_3})(\ref{BCD_FP_DCP_6_4})}.
	\end{aligned}
\end{equation}
By introducing $\gamma_{q,k}$ to the optimization objective, Problem $\mathcal P4d$ can be equivalently reformulated as
\begin{equation}
    \label{BCD_FP_DCP_9}
    \begin{aligned}
		\mathcal P4e:& \max \limits _{{\boldsymbol{\theta},\boldsymbol{\mathbf{F}},\boldsymbol{\alpha^*}}} \sum^K_{k=1}\sum^Q_{q=1} \frac{\alpha^*_{q,k}\left\Vert{\bar{\mathbf H}}_{q,k}\mathbf{F}_{q,k}\right\Vert^2}{\sum^{Q,K}_{i=1,j=1}\left\Vert{\bar{\mathbf H}}_{i,k}\mathbf{F}_{i,j}\right\Vert^2+\sigma^2}\\
		\text{s.t.}\ &\text{(\ref{BCD_FP_DCP_6_2})(\ref{BCD_FP_DCP_6_3})(\ref{BCD_FP_DCP_6_4})},
	\end{aligned}
\end{equation}
where ${\alpha^*_{q,k}=\omega^*_{q,k}(1+\alpha_{q,k})}$.

As the objective function in Problem $\mathcal P4e$ is in a fractional form, the quadratic fractional programming that introduces well-designed auxiliary variables \cite{shen2018fractional} is utilized. By decoupling the numerator and the denominator of each ratio term, the subproblem can be formulated as
\begin{equation}
    \label{BCD_FP_DCP_10}
    \begin{aligned}
		\mathcal P4f:& \max \limits _{\boldsymbol{\boldsymbol{\theta},\mathbf{F}}} \sum^K_{k=1}\sum^Q_{q=1} ( 2\rho_{q,k}\sqrt{\alpha^*_{q,k}}\Vert{\bar{\mathbf H}}_{q,k}\mathbf{F}_{q,k}\Vert\\
        &-\rho^2_{q,k}( \sum^K_{j=1}\Vert{\bar{\mathbf H}}_{q,j}\mathbf{F}_{q,j}\Vert^2+\sigma^2))\\
		\text{s.t.}\ &\text{(\ref{BCD_FP_DCP_6_2})(\ref{BCD_FP_DCP_6_3})(\ref{BCD_FP_DCP_6_4})},
	\end{aligned}
\end{equation}
where $\boldsymbol\rho$ is the auxiliary vector of the quadratic transform fractional programming. Based on the Lagrange multiplier method, the optimal $\rho_{q,k}$ is given by

\begin{equation}
	{\rho_{q,k}}=\frac{ \sqrt{\alpha^*_{q,k}}\Vert{\bar{\mathbf H}}_{q,k}\mathbf{F}_{q,k}\Vert           }{         {\sum^{Q,K}_{i=1,j=1}\left\Vert{\bar{\mathbf H}}_{i,k}\mathbf{F}_{i,j}\right\Vert^2+\sigma^2}      }.
	\label{rho_l_k}
\end{equation}

We introduce substituted decision variables $\Theta_n\triangleq e^{j\theta_n},\forall n$ to replace $\theta_n$ ($0 \leq \theta _{n} < 2\pi$) and thus $\Vert\Theta_n\Vert=1$ is in a nonconvex set. We need to relax the constraints as $\Vert\Theta_n\Vert<1$, when the decision variables $\Theta_n$ are in a convex set. The subproblem can be formulated as

\begin{subequations}
    \label{BCD_FP_DCP_11}
    \begin{align}
		\mathcal P4g:& \max \limits _{\boldsymbol{\boldsymbol{\Theta},\mathbf{F}}} \sum^K_{k=1}\sum^Q_{q=1} ( 2\rho_{q,k}\sqrt{\alpha^*_{q,k}}\Vert{\bar{\mathbf H}}_{q,k}\mathbf{F}_{q,k}\Vert\nonumber\\
        &-\rho^2_{q,k}( \sum^K_{j=1}\Vert{\bar{\mathbf H}}_{q,j}\mathbf{F}_{q,j}\Vert^2+\sigma^2))\nonumber\\
		\text{s.t.}\ & {\bar{\mathbf H}}_{q,k}=G_{q,R} \boldsymbol{\Phi} H_{R,k}^H +\mathbf{H}_{q,k},\label{BCD_FP_DCP_11_1}\\
		& \boldsymbol{\Phi}=diag\{\Theta_1,\Theta_2,...,\Theta_M\},\label{BCD_FP_DCP_11_2}\\
		& \Vert\Theta_n\Vert < 1, \forall n,\\
		& \Vert\mathbf{F}_{q,k}\Vert < 1, \forall q,k.
	\end{align}
\end{subequations}

Problem $\mathcal P4g$ is a difference-of-convex problem. For simplification, we define $h(x)$ $=\sum ^{Q,K}_{q=1,k=1}  2\rho_{q,k}\sqrt{\alpha^*_{q,k}}$ $\Vert{\bar{\mathbf H}}_{q,k}\mathbf{F}_{q,k}\Vert$ and $g(x)=-\sum ^{Q,K}_{q=1,k=1} \rho^2_{q,k}\left( \sum^K_{j=1}\Vert{\bar{\mathbf H}}_{q,j}\mathbf{F}_{q,j}\Vert^2+\sigma^2\right)$, where $x$ is the decision variables.

The difference-of-convex problem $\mathcal P4g$ can be solved by Majorization Minimization (MM) interpretation as
\begin{equation}
	\label{BCD_FP_DCP_12}
	x^{m+1}=\arg \min_{x} \{ g(x)+\nabla h(x^m)(x-x^m) \}
\end{equation}
where $arg \min$ is the point at which the function values are minimized and $\nabla$ is the gradient.

Controlling the phase shifts of the reflected signals at the IRS is referred to as the passive beamforming (PBF), while the precoding operation at the cell-edge users is termed as the active beamforming (ABF)\cite{sur2021sum}. Relying on the BCD method, the subproblem (\ref{BCD_FP_DCP_11}) is decoupled into two parts to optimize PBF and ABF settings alternatively.

For the passive beamforming optimization, the MM is utilized to solve the $m^{th}$ iteration, which corresponds to Problem $\mathcal P4h$ as follows.
\begin{subequations}
    \label{BCD_FP_DCP_13}
    \begin{align}
		\mathcal P4h:& \min \limits _{\boldsymbol{\boldsymbol{\theta}}} \sum^K_{k=1}\sum^Q_{q=1} \rho^2_{q,k}\left( \sum^K_{j=1}\Vert{\bar{\mathbf H}}_{q,j}\mathbf{F}_{q,j}\Vert^2+\sigma^2\right)\nonumber\\
        &+\nabla h(\boldsymbol{\theta}^m)(\boldsymbol{\theta}-\boldsymbol{\theta}^m)\nonumber\\
		\text{s.t.}\quad & \Vert\Theta_n\Vert < 1, \forall n,\\
		&\text{(\ref{BCD_FP_DCP_11_1})(\ref{BCD_FP_DCP_11_2})}\nonumber.
	\end{align}
\end{subequations}
In each iteration, Problem $\mathcal P4h$ is a convex problem and can be solved by CVX. For the active beamforming optimization, we use the MM to solve the $m^{th}$ iteration Problem $\mathcal P4i$ as follows.
\begin{subequations}
    \label{BCD_FP_DCP_14}
    \begin{align}
		\mathcal P4i:& \min \limits _{\boldsymbol{\mathbf{F}}} \sum^K_{k=1}\sum^Q_{q=1} \rho^2_{q,k}\left( \sum^K_{j=1}\Vert{\bar{\mathbf H}}_{q,j}\mathbf{F}_{q,j}\Vert^2+\sigma^2\right)\nonumber\\
        &+\nabla h(\boldsymbol{\mathbf{F}}^m)(\boldsymbol{\mathbf{F}}-\boldsymbol{\mathbf{F}}^m)\nonumber\\
		\text{s.t.}\ & \Vert\mathbf{F}_{q,k}\Vert < 1, \forall q,k,\\
		&\text{(\ref{BCD_FP_DCP_11_1})(\ref{BCD_FP_DCP_11_2})}\nonumber.
	\end{align}
\end{subequations}
In each iteration, Problem $\mathcal P4i$ is a convex problem and can be solved by CVX.

\subsection{Summary and Complexity}

\begin{algorithm}[t]
	\caption{BCD-FP-DC Algorithm}
	\begin{algorithmic}[1]
		\ENSURE ${\mathbf H}_{q,k},{\mathbf G}_{q,R},{\mathbf H}_{R,k}, L_{k}, f^{\mathbb E}_{q,\text{total}}, \omega_{k}, \zeta, B, \sigma^2$
		\REQUIRE $\mathbf{F},\boldsymbol{\theta},\boldsymbol\ell,\boldsymbol f^{\mathbb E}$
		\STATE $n=1$, Calculate ${Cost}^n=\sum ^{K}_{k=1} \omega_{k} (E_k+\zeta D_k)$ by (\ref{cost_function}).
		\STATE Initialize all the optimization variables $\mathbf{F},\boldsymbol{\theta},\boldsymbol\ell,\boldsymbol f^{\mathbb E}$ with random values.
		\WHILE {${Cost}^n-{Cost}^{n-1}>\epsilon$, \AND $n<N$}
		    \STATE Update $n=n+1$.
			\STATE Fix $\mathbf{F}^n$, $\boldsymbol{\theta}^n$, and update $\boldsymbol\ell^{n+1}$, ${\boldsymbol f^{\mathbb E}}^{n+1}$ by solving the MEC subproblem (\ref{BCD_FP_DCP_2}) using Gurobi\cite{gurobi2018gurobi}.
			\STATE Fix $\boldsymbol\ell^{n+1}$ and ${\boldsymbol f^{\mathbb E}}^{n+1}$ to solve the IRS communication subproblem:
			\STATE Update auxiliary variable $\beta$ and $\lambda$ using the modified Newton’s method in \cite{bai2020latency}. 
			\STATE Update auxiliary weight $\omega^*$ by $\omega_{q,k}^*=\lambda_{q,k}\beta_{q,k}$. 
			\STATE Update FP parameters. Update $\alpha_{q,k}=\gamma_{q,k}$ by (\ref{BCD_FP_DCP_7_1}). Update $\alpha^*_{q,k}$ by ${\alpha^*_{q,k}=\omega^*_{q,k}(1+\alpha_{q,k})}$. Update $\rho_{q,k}$ by (\ref{rho_l_k}).
			\STATE Update $m=1$.
			\REPEAT
			    \STATE Update $m=m+1$, and $\nabla h(\boldsymbol{\theta}^m)$.
				\STATE Solve the Majorization Minimization interpretation (\ref{BCD_FP_DCP_13}) by CVX and get $\boldsymbol{\theta}^{m+1}$.
			\UNTIL{Convergence}
			\STATE Update $\boldsymbol{\theta}^{n+1}$.
			\STATE Update $m=1$.
			\REPEAT
			    \STATE Update $m=m+1$, $\nabla h(\boldsymbol{\mathbf{F}}^m)$.
				\STATE Solve the Majorization Minimization interpretation (\ref{BCD_FP_DCP_14}) by CVX and get $\boldsymbol{\mathbf{F}}^{m+1}$.
			\UNTIL{Convergence}
			\STATE Update $\mathbf{F}^{n+1}$.
			\STATE Calculate ${Cost}^n=\sum ^{K}_{k=1} \omega_{k} (E_k+\zeta D_k)$ by (\ref{cost_function}).
		\ENDWHILE
	\end{algorithmic}
	\label{BCD_FP_DCP_algorithm}
\end{algorithm}

The detailed procedure of the proposed algorithm named as BCD-FP-DC is summarized as the pseudo-code in Algorithm \ref{BCD_FP_DCP_algorithm}. The outer loop of the BCD-FP-DC algorithm has the complexity of $\mathcal{O}(N_{BCD})$, where $N_{BCD}$ is the number of iterations of the BCD algorithm. The number of operations of the Gurobi's spatial branch-and-bound algorithm for solving the non-convex MEC subproblem is $N_{SBB}$ and the complexity of it is $\mathcal{O}(N_{SBB})$. $N_{COM}$ is the number of iterations that we need to perform CVX algorithm. We assume that the computation complexity of CVX program is $\mathcal{O}(M_{cvx})$. Let $N_v$ denote the number of variables. The CVX program we used is based on the interior-point method, which is a widely used optimization algorithm for solving convex optimization problems, where $M_{cvx}=N_v^{3.5}$\cite{grant2014cvx,gurobi2018gurobi}. Therefore, the total computational complexity of BCD-FP-DC is $\mathcal{O}(\max\{N_{BCD}N_{SBB},N_{BCD}N_{COM}N_v^{3.5}\})$.  In our proposed algorithm, the number of optimization variables involved in the IRS communication subproblem is significantly larger than that in the MEC subproblem, because the number of IRS elements typically far exceeds the number of users. Consequently, the computational complexity of solving the IRS subproblem is generally much higher than that of the MEC subproblem, i.e., $N_{COM}N_v^{3.5}>>N_{SBB}$. Therefore, the complexity of our BCD-FP-DC algorithm can be expressed as $\mathcal{O}(N_{BCD}N_{COM}N_v^{3.5})$.

\section{Numerical Results and Comparison}

In this section, the performance of the proposed algorithm is evaluated. The coordinates of two BSs are $(10m,-100m,0)$ and $(10m,100m,0)$, respectively. The number of elements of IRS is $N$, and the coordinate of IRS is $(-10m,0,1m)$. The number of antennas of BSs and cell-edge users are 3 and 2, respectively. The $K$ cell-edge users are located randomly at the edge of the cells. The detailed simulation settings are shown in Table \ref{Simulation_settings}.

\begin{table}[t]
\renewcommand{\arraystretch}{1.3}
\caption{Simulation settings}
\label{Simulation_settings}
\centering
\begin{tabular}{c|c||c|c}
\hline
\bfseries Parameter & \bfseries Value & \bfseries Parameter & \bfseries Value\\
\hline
Carrier freq.\cite{dampahalage2020intelligent} & 2.005GHz &  Bandwidth \cite{dampahalage2020intelligent} & 1kHz\\
Num. of BSs & 2 &  Num. of users & K\\
BS 1 location & (10m,-100m,0) & BS 2 location & (10m,100m,0)\\
IRS elements & N & IRS location & (-10m,0,1m)\\
BS antennas & 3 &  User antennas & 2\\
Noise \cite{dampahalage2020intelligent} & $3.16\times 10^{-11}$ &  $f^{\mathbb E}_{q,\text{total}}$ & 100 cycles/s\\
$L_{k}$ & 1000 & & \\
\hline
\end{tabular}
\end{table}

The performance of the proposed BCD-FP-DC algorithm is compared with three different benchmarks as follows.
\begin{enumerate}
\item SA:
Simulated annealing (SA) is a metaheuristic to approximate global optimization in an ample search space to solve such nonconvex problems for the formulated multi-cell IRS-aided MEC optimization problem. The complexity is $\mathcal{O}(T_{SA}N_{SA}C_{SA})$, where $T_{SA}$ is the number of temperature levels in the cooling schedule, $N_{SA}$ is the average number of iterations performed at each temperature level, and $C_{SA}$ is the computational cost of evaluating the objective function and generating neighbors at each iteration.

\item BCD-SA:
BCD-assisted simulated annealing (BCD-SA) algorithm introduces SA to solve the complicated IRS communication subproblem. The MEC subproblem is solved by Gurobi \cite{gurobi2018gurobi}, similar to the proposed BCD-FP-DC algorithm. The complexity is $\mathcal{O}(N_{BCD}T_{SA2}N_{SA2}C_{SA2})$, where $T_{SA2}$ is the number of temperature levels in the cooling schedule, $N_{SA2}$ is the average number of iterations performed at each temperature level, and $C_{SA2}$ is the computational cost of evaluating the objective function and generating neighbors at each iteration.

\item BCD-MSE:
BCD-aided mean-square error (BCD-MSE) uses the BCD structure to decompose the problem into the MEC subproblem and the IRS communication subproblem. The mean-square error (MSE) \cite{pan2020multicell} method solves the IRS communication subproblem. BCD-MSE introduces linear decoding matrix ${\mathbf U}_{q,k}$, MSE matrix ${\mathbf E}_{q,k}$, and auxiliary matrix ${\mathbf W}_{q,k}= {\mathbf E}_{q,k}^{-1}$. The problem is transformed as
\begin{equation}
\begin{aligned}
	\max \limits _{{{\mathbf{W}}, {\mathbf{U}}, {\mathbf{F}},{ \boldsymbol {\theta }} } } &\sum \limits _{q = 1}^{Q} {\sum \limits _{k = 1}^{K} {{\omega _{q,k}}{h_{q,k}}\left ({{{\mathbf{W}}, {\mathbf{U}}, {\mathbf{F}},{ \boldsymbol {\theta }} } }\right)} }\\
	\textrm {s.t.}\quad & {h_{q,k}}={\log \left |{ {{{\mathbf{W}}_{q,k}}} }\right | - {\mathrm{Tr}}\left [{{{{\mathbf{W}}_{q,k}}{{\mathbf{E}}_{q,k}}} }\right] + d},\\
	& \Vert\mathbf{F}_{q,k}\Vert < 1, \forall q,k,\\
	& 0 \leq \theta_{n} < 2\pi, \forall n.
\end{aligned}
\label{p_2}
\end{equation}
where ${\mathbf U}_{q,k}$ can be updated by:
\begin{equation}
{\mathbf{U}}_{q,k} = {\left ({{\mathbf{J}}_{q,k}} + {{\bar {\mathbf  H}}_{q,k}}{{\mathbf{F}}_{q,k}}{\mathbf{F}}_{q,k}^{\mathrm{H}}\bar {\mathbf {H}}_{q,k}^{\mathrm{H}}\right)^{ - 1}}{\bar {\mathbf  H}}_{q,k}{{\mathbf{F}}_{q,k}}.
\label{U_l_k}
\end{equation}
The complexity is 
\begin{equation}
\mathcal{O}(N_{BCD}\max\{QKN_{BS}^3,L^2N_{BS}^2N_{U},C_{MM,CCM}\}),
\end{equation}
where $C_{MM,CCM}$ is the complexity of MM and complex circle manifold (CCM) algorithm in \cite{pan2020multicell}.

\end{enumerate}

Moreover, two extra benchmarks are compared:
\begin{enumerate}
    \item Rand-Phase: The phase shifts of IRS elements are uniformly and independently distributed in $[0,2\pi]$. Our proposed BCD-FP-DC algorithm is performed to solve the optimization problem.
    \item No-IRS: The IRS is removed from the system, i.e., ${\mathbf G}_{l,R}=0$, and ${\mathbf H}_{R,k}=0$. Our proposed BCD-FP-DC algorithm is used to solve the optimization problem.
\end{enumerate}

\subsection{Benchmarks Comparison}

\subsubsection{Convergence Behavior of BCD-FP-DC and Benchmarks}

\begin{figure}[t]
    \centering
    \includegraphics[width=0.47\textwidth]{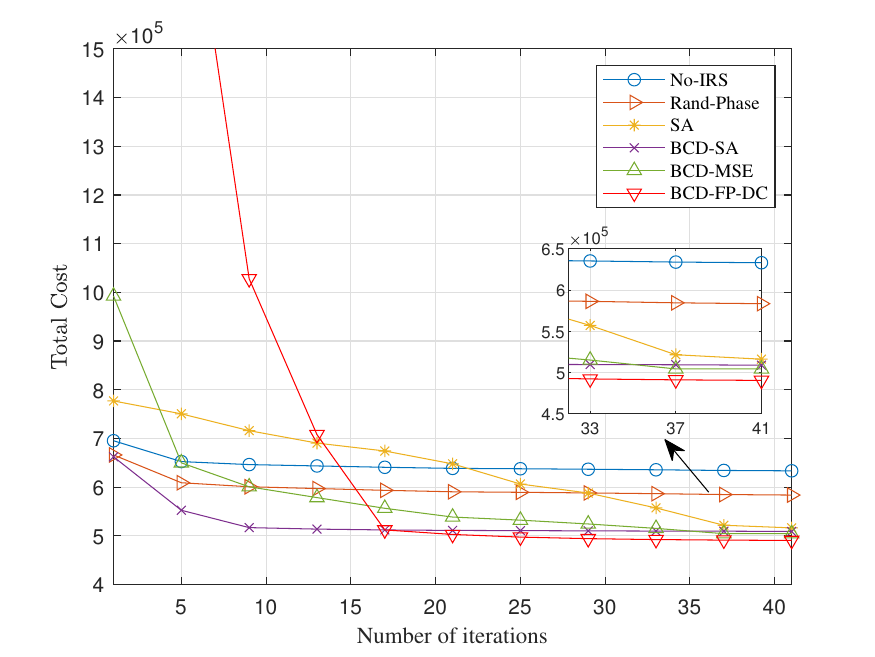}
    \caption{Convergence behaviour of proposed BCD-FP-DC algorithm and benchmarks with 64 IRS elements}
    \label{compare6}
\end{figure}

In Fig. \ref{compare6}, the IRS has 64 elements, and the number of cell-edge users is 3. Total cost is the optimization objective $\sum ^{K}_{k=1} \omega_{k} C_{k}$. The convergence performance of the proposed BCD-FP-DC algorithm is compared with different benchmarks, as shown in Fig. \ref{compare6}. It can be observed that the average system cost converges rapidly within 40 iterations. In Fig. \ref{compare6}, our proposed BCD-FP-DC algorithm outperforms two heuristic methods and the BCD-MSE algorithm. 
Especially as can be seen from Fig. \ref{compare6}, BCD can be treated as a promising technique to achieve smaller total cost when it comes to converging. Moreover, the proposed algorithm BCD-FP-DC needs around 20 iterations to converge to a lower system cost due to more iterations being required to reach a suitable auxiliary variable to transform the original problem into a weight-sum-rate problem. The proposed algorithm and benchmark algorithms outperform the scenarios without IRS or with random phases in IRS, and this proves that the multi-cell IRS-aided MEC system achieves further performance gain compared to the system without IRS.

\subsubsection{Impact of Different Numbers of IRS Elements of BCD-FP-DC and Benchmarks}

\begin{figure}[t]
    \centering
    \includegraphics[width=0.47\textwidth]{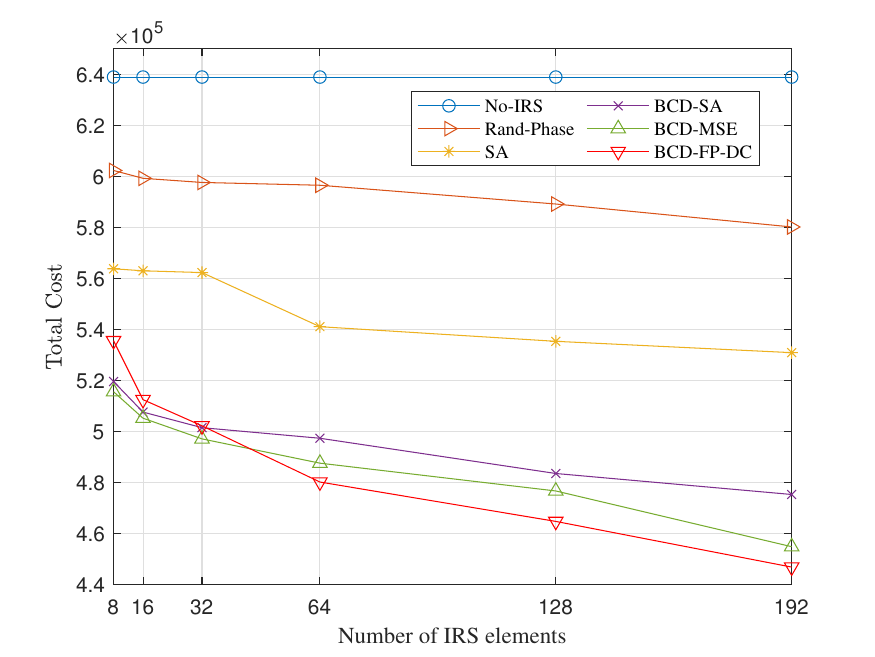}
    \caption{The total cost versus different numbers of IRS elements}
    \label{Cost_algor_IRS}
\end{figure}

Fig. \ref{Cost_algor_IRS} shows the total cost verse the different number of IRS elements among the proposed algorithm and the benchmarks.
It can be observed from the numerical results that the proposed algorithm and benchmark algorithms outperform the scenario under no or random IRS.
Our proposed BCD-FP-DC algorithm achieves more cost reduction under large-scale IRS elements compared to the three benchmark algorithms. Both the BCD-FP-DC algorithm and BCD-MSE algorithm have similar performance under different numbers of IRS elements, and BCD-FP-DC slightly outperforms BCD-MSE. Additionally, the total cost reduction becomes significant when increasing the number of IRS elements because more IRS elements enhance the received signal power and provide a high achievable data rate to save energy and reduce latency.

\subsubsection{Impact of Different Numbers of Cell-edge Users of BCD-FP-DC and Benchmarks}

\begin{figure}[t]
    \centering
    \includegraphics[width=0.47\textwidth]{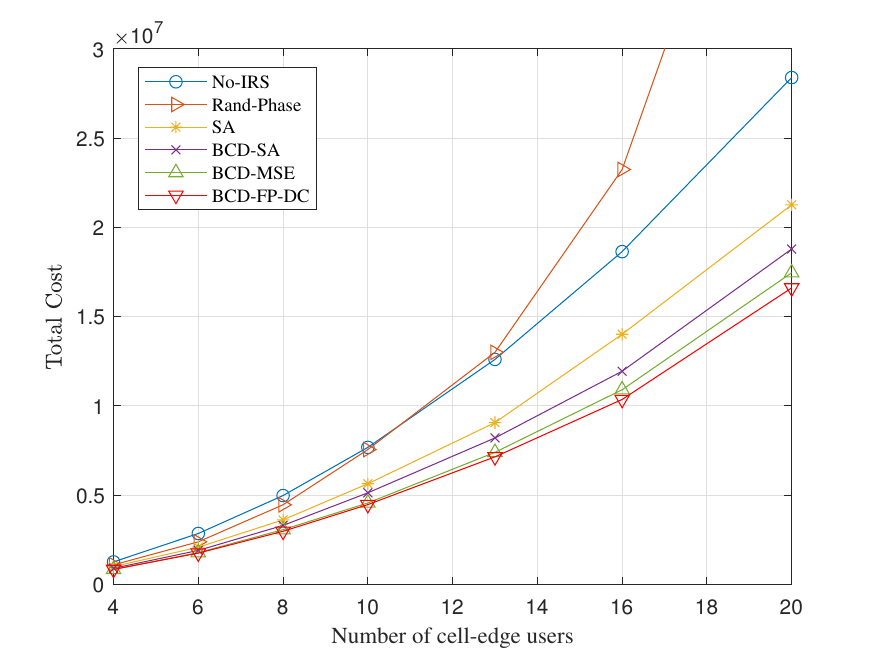}
    \caption{The total cost versus different numbers of cell-edge users with 64 IRS elements}
    \label{Cost_algor_users}
\end{figure}

As shown in Fig. \ref{Cost_algor_users}, the total cost increases rapidly with the growing number of cell-edge users. Both BCD-FP-DC and BCD-MSE have similar performance over the entire range of cell-edge users. The random-phase IRS can enhance the system performance compared to the scheme without using IRS when there are few mobile cell-edge users. However, as the number of users increases, the random-phase IRS may cause interference, leading to performance degradation. Our proposed BCD-FP-DC algorithm outperforms all the benchmarks with different numbers of cell-edge users. The IRS can help improve the performance of multi-cell MEC because all the algorithms (SA, BCD-SA, BCD-MSE, BCD-FP-DC) with IRS outperform the multi-cell MEC system without the help of the IRS.

\subsubsection{Impact of Different Numbers of BS Antennas of BCD-FP-DC and Benchmarks}

\begin{figure}[t]
    \centering
    \includegraphics[width=0.47\textwidth]{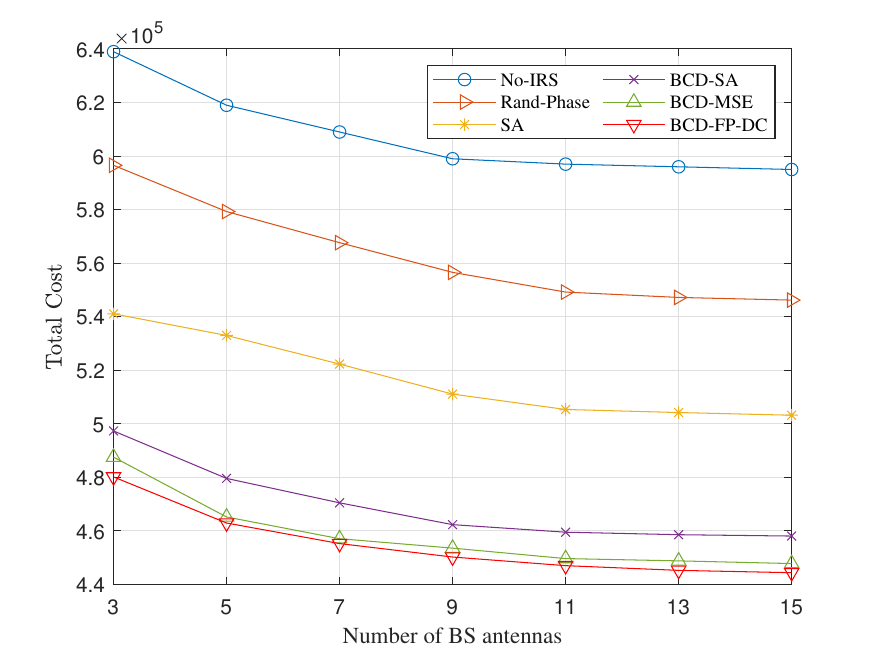}
    \caption{The total cost versus different numbers of BS antennas with 64 IRS elements}
    \label{Cost_algor_BSAntennas}
\end{figure}

Fig. \ref{Cost_algor_BSAntennas} shows the total cost verse the different number of BS antennas among the proposed algorithm and the benchmarks. Similar to increasing the number of BS antennas in MIMO systems can improve communication efficiency, in our proposed system, increasing the number of BS antennas can also reduce the optimization objective of the total cost. More antennas can help decrease the total cost by reducing propagation delay and lowering energy consumption during transmission.

\subsubsection{Complexity Comparison}

\begin{table}[t]
\renewcommand{\arraystretch}{1.3}
\caption{Complexity Comparison}
\label{Complexity}
\centering
\begin{tabular}{c|c}
\hline
\bfseries Algorithm & \bfseries Complexity\\
\hline
BCD-FP-DC & $\mathcal{O}(N_{BCD}N_{COM}N_v^{3.5})$\\
SA & $\mathcal{O}(T_{SA}N_{SA}C_{SA})$\\
BCD-SA & $\mathcal{O}(N_{BCD}T_{SA2}N_{SA2}C_{SA2})$\\
BCD-MSE & $\mathcal{O}(N_{BCD}\max\{QKN_{BS}^3,L^2N_{BS}^2N_{U},C_{MM,CCM}\})$\\
Rand-Phase & $\mathcal{O}(N_{BCD}N_{SBB})$\\
No-IRS & $\mathcal{O}(N_{BCD}N_{SBB})$\\
\hline
\end{tabular}
\end{table}

\begin{figure}[t]
\centering
\includegraphics[width=0.47\textwidth]{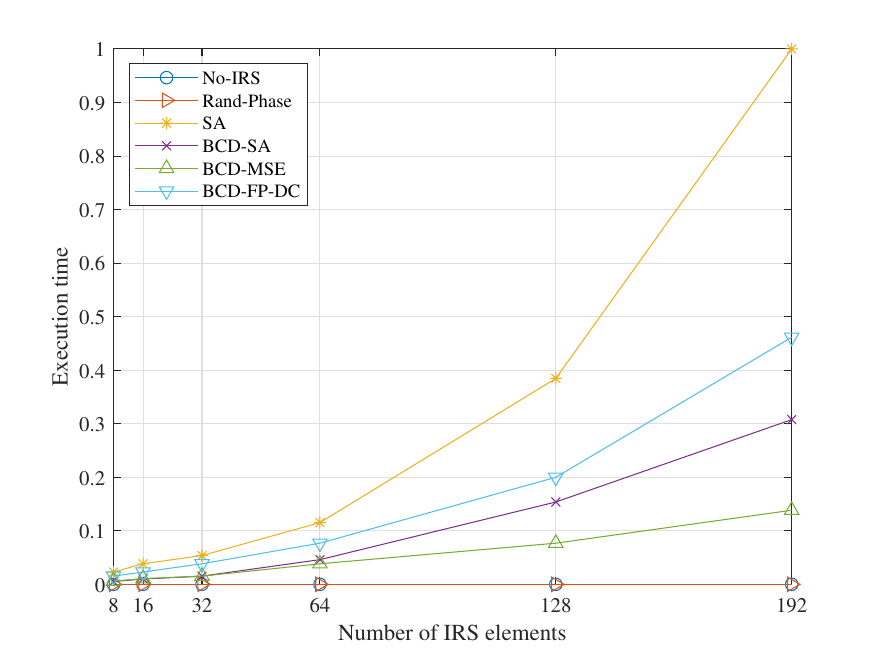}
\caption{The execution time against different numbers of IRS elements}
\label{complexity_fig}
\end{figure}

We present the computational complexities of the proposed algorithm and the benchmark algorithms, as shown in Table \ref{Complexity}. A direct comparison of complexities can be challenging; therefore, we provide normalized algorithm execution times as a comparative metric in Fig. \ref{complexity_fig}, with all algorithms running until convergence. No-IRS and Rand-Phase exhibit the best performance across all numbers of IRS elements, which indicates that the controller spends the least time generating optimization results when not required to determine the phase shift. SA has the worst performance on execution time among all algorithms, especially at larger numbers of IRS elements. In comparison with the BCD-MSE and BCD-SA algorithms, although our proposed BCD-FP-DC algorithm has higher complexity, it possesses advantages in reducing the total cost of the system.

\subsection{Analysis of the Proposed BCD-FP-DC Algorithm}

\subsubsection{Convergence Behavior of BCD-FP-DC with Different Number of IRS Elements}

\begin{figure}[t]
\centering
\includegraphics[width=0.47\textwidth]{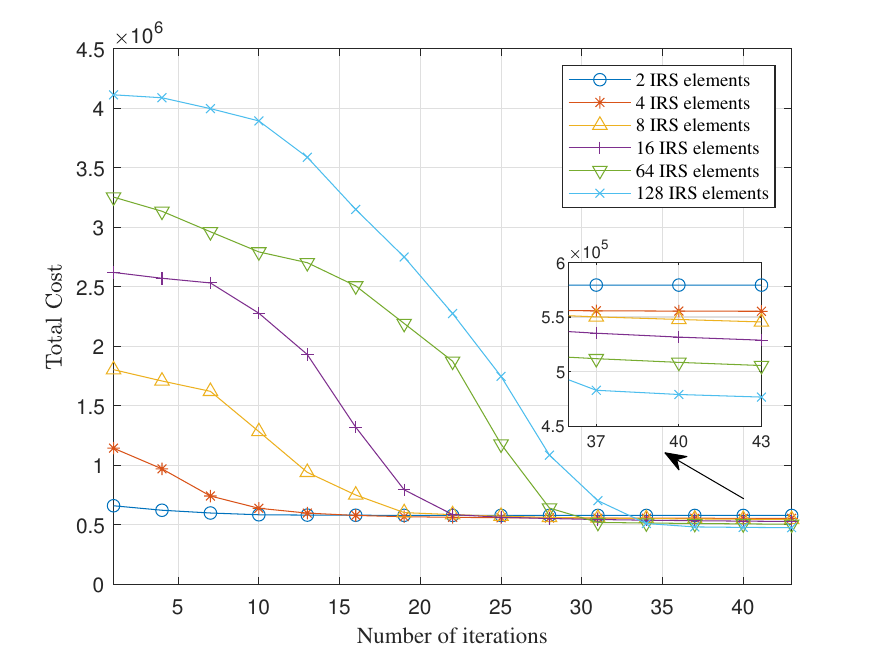}
\caption{Convergence behaviour of proposed BCD-FP-DC algorithm with different numbers of IRS elements}
\label{fraction_result4}
\end{figure}

We introduce greater ambient noise with variance as $\sigma^2=3.16\times 10^{-9}$ during the simulation to slow down the convergence to better illustrate the convergence behavior. The numerical results in Fig. \ref{fraction_result4} show that fewer IRS elements result in higher convergence speed. Furthermore, the total cost of the proposed BCD-FP-DC algorithm decreases with the increasing number of IRS elements.

\subsubsection{Energy and Latency Analysis of BCD-FP-DC}

\begin{figure}[t]
\centering
\includegraphics[width=0.47\textwidth]{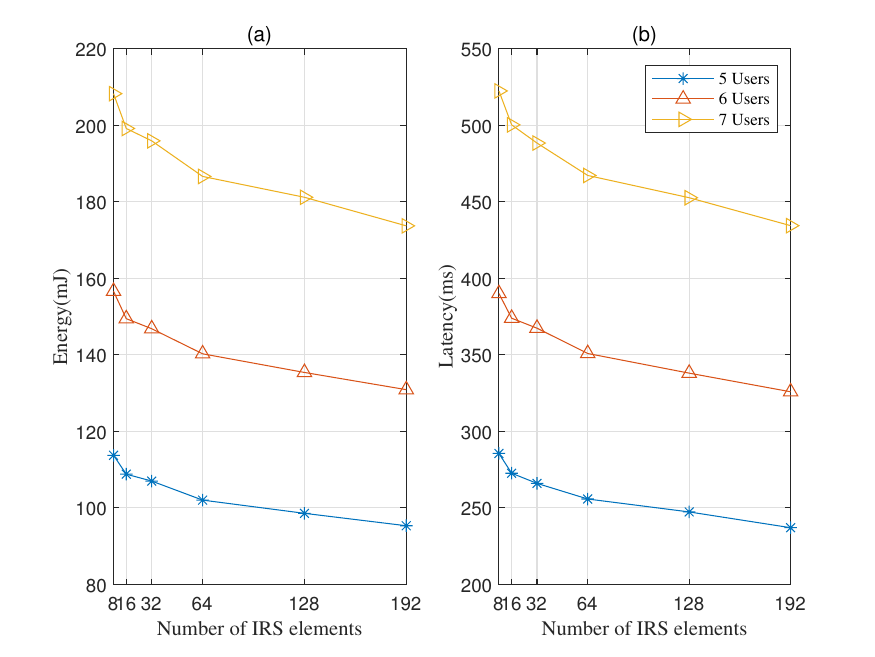}
\caption{The energy and the latency performance of the proposed BCD-FP-DC algorithm verse different numbers of IRS elements with different number of cell-edge users}
\label{fraction_4costs}
\end{figure}

\begin{figure}[t]
\centering
\includegraphics[width=0.47\textwidth]{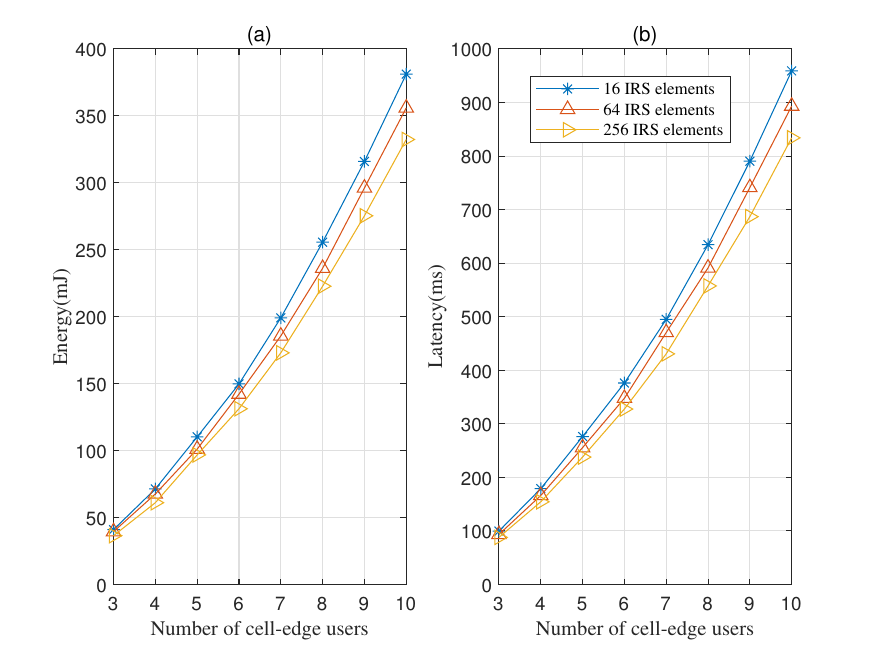}
\caption{The energy and the latency performance of the proposed BCD-FP-DC algorithm verse different numbers of cell-edge users with different number of IRS elements}
\label{fraction_5costs}
\end{figure}

Fig. \ref{fraction_4costs} and Fig. \ref{fraction_5costs} show the curve of execution latency and energy consumption separately. Fig. \ref{fraction_4costs} shows that both execution latency and energy consumption tend to decrease with the growing number of IRS elements, validating that more IRS elements will improve both of them. Fig. \ref{fraction_5costs} shows that both execution latency and energy consumption tend to increase with the growing number of cell-edge users, validating that more users will introduce more interference and degrade the system performance.

\section{Conclusion}

In this paper, we proposed a multi-cell IRS-aided MEC scheme where IRS can resolve the link blockage problems to guarantee the offloading efficiency in multi-cell networks. We aimed to minimize the joint energy and latency cost by jointly optimizing the MEC and IRS communication settings while satisfying the system constraints. The formulated problem was decomposed by BCD into two subproblems, i.e., the MEC subproblem and the IRS communication subproblem. We then proposed the BCD-FP-DC algorithm that can alternatively optimize the MEC resources and the IRS communication until convergence. Numerical results validated that our proposed algorithm can outperform all the benchmarks under large-scale IRS elements. Moreover, the multi-cell IRS-aided MEC framework can achieve further performance gains compared to the multi-cell MEC system without the help of IRS. In our future work, we will further investigate online algorithms that adapt to varying wireless environments in multi-cell IRS-aided MEC.

\appendices
\section{Proof of the reformulation from Problem $\mathcal P3a$ to Problem $\mathcal P3b$}
\label{P3a_P3b}
In order to elucidate the transformation process from problem $\mathcal P3a$ to problem $\mathcal P3b$, we express each new variable in problem $\mathcal P3b$ as a representation of the variables present in problem $\mathcal P3a$. The specific values of $d$, $i$, $m$, $s_i$, $c^T$, $Q_i$ and $r_i^T$ in $\mathcal P3b$ are:
\begin{subequations}
    \label{value_of_p3b}
	\begin{align}
        &d=0,\quad i=(q-1)K+k,\quad m=QK, \quad s_i=0,\\
        &c^T\boldsymbol{x}=\sum ^{K}_{k=1} \omega_{k} (E_k+\zeta D_k),\\
		&\frac{1}{2} \boldsymbol{x}^TQ_i\boldsymbol{x}=\frac{1}{BR_{q,k}}\ell _{q,k}f^{\mathbb E}_{q,k}-D_kf^{\mathbb E}_{q,k},\\
        &r_i^T\boldsymbol{x}=\ell _{q,k}c_k,
	\end{align}
\end{subequations}
where $c=[\underbrace {0,\cdots,0}_{\mathrm{QK~zeros}},\omega_1,\cdots,\omega_K,\omega_1\zeta,\cdots,\omega_K\zeta,\underbrace {0,\cdots,0}_{\mathrm{QK~zeros}}]^T$, $Q_i\in R^{2(Q+1)K\times 2(Q+1)K}$ is a sparse matrix with only four non-zero entries at positions $\{(q-1)K+k,(Q+1+q)K+k\}$, $\{(Q+1+q)K+k),(q-1)K+k\}$, $\{(Q+1)K+k,(Q+1+q)K+k\}$, $\{(Q+1+q)K+k,(Q+1)K+k\}$, with corresponding values $\frac{1}{BR_{q,k}}$, $\frac{1}{BR_{q,k}}$, $-1$, $-1$, respectively, and $r_i$ is a vector with only one non-zero element $c_k$ at the $i^{th}$ entry.





\ifCLASSOPTIONcaptionsoff
  \newpage
\fi



\bibliographystyle{IEEEtran}
\bibliography{IEEEabrv,bare_jrnl}

\begin{thebibliography}{10}
\providecommand{\url}[1]{#1}
\csname url@samestyle\endcsname
\providecommand{\newblock}{\relax}
\providecommand{\bibinfo}[2]{#2}
\providecommand{\BIBentrySTDinterwordspacing}{\spaceskip=0pt\relax}
\providecommand{\BIBentryALTinterwordstretchfactor}{4}
\providecommand{\BIBentryALTinterwordspacing}{\spaceskip=\fontdimen2\font plus
\BIBentryALTinterwordstretchfactor\fontdimen3\font minus \fontdimen4\font\relax}
\providecommand{\BIBforeignlanguage}[2]{{%
\expandafter\ifx\csname l@#1\endcsname\relax
\typeout{** WARNING: IEEEtran.bst: No hyphenation pattern has been}%
\typeout{** loaded for the language `#1'. Using the pattern for}%
\typeout{** the default language instead.}%
\else
\language=\csname l@#1\endcsname
\fi
#2}}
\providecommand{\BIBdecl}{\relax}
\BIBdecl

\bibitem{mao2017survey}
Y.~Mao, C.~You, J.~Zhang, K.~Huang, and K.~B. Letaief, ``A survey on mobile edge computing: The communication perspective,'' \emph{{IEEE} Commun. Surveys Tuts.}, vol.~19, no.~4, pp. 2322--2358, 2017.

\bibitem{sun2016edgeiot}
X.~Sun and N.~Ansari, ``{EdgeIoT}: Mobile edge computing for the internet of things,'' \emph{IEEE Commun. Mag.}, vol.~54, no.~12, pp. 22--29, 2016.

\bibitem{you2016energy}
C.~You, K.~Huang, H.~Chae, and B.-H. Kim, ``Energy-efficient resource allocation for mobile-edge computation offloading,'' \emph{IEEE Trans. Wireless Commun.}, vol.~16, no.~3, pp. 1397--1411, 2016.

\bibitem{computing2014mobile}
M.-E. Computing, I.~Initiative \emph{et~al.}, ``Mobile-edge computing,'' \emph{Introductory Technical White Paper}, 2014.

\bibitem{hu2015mobile}
Y.~C. Hu, M.~Patel, D.~Sabella, N.~Sprecher, and V.~Young, ``Mobile edge computing—a key technology towards {5G},'' \emph{ETSI white paper}, vol.~11, no.~11, pp. 1--16, 2015.

\bibitem{yang2022intelligent}
Y.~Yang, Y.~Gong, and Y.-C. Wu, ``Intelligent reflecting surface aided mobile edge computing with binary offloading: Energy minimization for {IoT} devices,'' \emph{IEEE Internet Things J.}, vol.~9, no.~15, pp. 12\,973--12\,983, 2022.

\bibitem{mach2017mobile}
P.~Mach and Z.~Becvar, ``Mobile edge computing: A survey on architecture and computation offloading,'' \emph{IEEE Commun. Surveys Tuts.}, vol.~19, no.~3, pp. 1628--1656, 2017.

\bibitem{wu2018noma}
Y.~Wu, K.~Ni, C.~Zhang, L.~P. Qian, and D.~H. Tsang, ``Noma-assisted multi-access mobile edge computing: A joint optimization of computation offloading and time allocation,'' \emph{IEEE Trans. Veh. Technol.}, vol.~67, no.~12, pp. 12\,244--12\,258, 2018.

\bibitem{chu2020intelligent}
Z.~Chu, P.~Xiao, M.~Shojafar, D.~Mi, J.~Mao, and W.~Hao, ``Intelligent reflecting surface assisted mobile edge computing for internet of things,'' \emph{IEEE Wireless Commun. Lett.}, vol.~10, no.~3, pp. 619--623, 2020.

\bibitem{yu2022deep}
J.~Yu, X.~Liu, Y.~Gao, C.~Zhang, and W.~Zhang, ``Deep learning for channel tracking in {IRS}-assisted {UAV} communication systems,'' \emph{IEEE Trans. Wireless Commun.}, vol.~21, no.~9, pp. 7711--7722, 2022.

\bibitem{shao2022target}
X.~Shao, C.~You, W.~Ma, X.~Chen, and R.~Zhang, ``Target sensing with intelligent reflecting surface: Architecture and performance,'' \emph{IEEE J. Sel. Areas Commun.}, vol.~40, no.~7, pp. 2070--2084, 2022.

\bibitem{shi2022intelligent}
W.~Shi, W.~Xu, X.~You, C.~Zhao, and K.~Wei, ``Intelligent reflection enabling technologies for integrated and green internet-of-everything beyond {5G}: Communication, sensing, and security,'' \emph{IEEE Wireless Commun.}, vol.~30, no.~2, pp. 147--154, 2023.

\bibitem{yu2021irs}
X.~Yu, D.~Xu, D.~W.~K. Ng, and R.~Schober, ``{IRS}-assisted green communication systems: Provable convergence and robust optimization,'' \emph{IEEE Trans. Commun.}, vol.~69, no.~9, pp. 6313--6329, 2021.

\bibitem{pan2022sum}
Y.~Pan, K.~Wang, C.~Pan, H.~Zhu, and J.~Wang, ``Sum-rate maximization for intelligent reflecting surface assisted terahertz communications,'' \emph{IEEE Trans. Veh. Technol.}, vol.~71, no.~3, pp. 3320--3325, 2022.

\bibitem{wu2021intelligent}
Q.~Wu, X.~Guan, and R.~Zhang, ``Intelligent reflecting surface-aided wireless energy and information transmission: An overview,'' \emph{Proc. IEEE}, vol. 110, no.~1, pp. 150--170, 2022.

\bibitem{ge20165g}
X.~Ge, S.~Tu, G.~Mao, C.-X. Wang, and T.~Han, ``{5G} ultra-dense cellular networks,'' \emph{IEEE Wireless Commun.}, vol.~23, no.~1, pp. 72--79, 2016.

\bibitem{chen2022dynamic}
X.~Chen, Y.~Bi, X.~Chen, H.~Zhao, N.~Cheng, F.~Li, and W.~Cheng, ``Dynamic service migration and request routing for microservice in multi-cell mobile edge computing,'' \emph{IEEE Internet Things J.}, vol.~9, no.~15, pp. 13\,126--13\,143, 2022.

\bibitem{hua2019reconfigurable}
S.~Hua and Y.~Shi, ``Reconfigurable intelligent surface for green edge inference in machine learning,'' in \emph{GC Wkshps}.\hskip 1em plus 0.5em minus 0.4em\relax IEEE, 2019, pp. 1--6.

\bibitem{cao2019intelligent}
Y.~Cao and T.~Lv, ``Intelligent reflecting surface enhanced resilient design for {MEC} offloading over millimeter wave links,'' \emph{arXiv preprint arXiv:1912.06361}, 2019.

\bibitem{huang2021reconfigurable}
S.~Huang, S.~Wang, R.~Wang, M.~Wen, and K.~Huang, ``Reconfigurable intelligent surface assisted edge machine learning,'' in \emph{ICC 2021-IEEE Int. Conf. Commun.}\hskip 1em plus 0.5em minus 0.4em\relax IEEE, 2021, pp. 1--6.

\bibitem{huang2021reconfigurable2}
------, ``Reconfigurable intelligent surface assisted mobile edge computing with heterogeneous learning tasks,'' \emph{IEEE Trans. Cogn. Commun. Netw.}, vol.~7, no.~2, pp. 369--382, 2021.

\bibitem{bai2020latency}
T.~Bai, C.~Pan, Y.~Deng, M.~Elkashlan, A.~Nallanathan, and L.~Hanzo, ``Latency minimization for intelligent reflecting surface aided mobile edge computing,'' \emph{IEEE J. Sel. Areas Commun.}, vol.~38, no.~11, pp. 2666--2682, 2020.

\bibitem{liu2021latency}
Y.~Liu, Q.~Hu, Y.~Cai, and M.~Juntti, ``Latency minimization in intelligent reflecting surface assisted {D2D} offloading systems,'' \emph{IEEE Commun. Lett.}, vol.~25, no.~9, pp. 3046--3050, 2021.

\bibitem{zhou2020delay}
F.~Zhou, C.~You, and R.~Zhang, ``Delay-optimal scheduling for {IRS}-aided mobile edge computing,'' \emph{IEEE Wireless Commun. Lett.}, vol.~10, no.~4, pp. 740--744, 2020.

\bibitem{li2021energy}
Z.~Li, M.~Chen, Z.~Yang, J.~Zhao, Y.~Wang, J.~Shi, and C.~Huang, ``Energy efficient reconfigurable intelligent surface enabled mobile edge computing networks with {NOMA},'' \emph{IEEE Trans. Cogn. Commun. Netw.}, vol.~7, no.~2, pp. 427--440, 2021.

\bibitem{huang2022integrated}
N.~Huang, T.~Wang, Y.~Wu, Q.~Wu, and T.~Q. Quek, ``Integrated sensing and communication assisted mobile edge computing: An energy-efficient design via intelligent reflecting surface,'' \emph{IEEE Wireless Commun. Lett.}, vol.~11, no.~10, pp. 2085--2089, 2022.

\bibitem{xu2022energy}
Z.~Xu, J.~Liu, J.~Zou, and Z.~Wen, ``Energy-efficient design for {IRS}-assisted {NOMA}-based mobile edge computing,'' \emph{IEEE Commun. Lett.}, vol.~26, no.~7, pp. 1618--1622, 2022.

\bibitem{hua2021reconfigurable}
S.~Hua, Y.~Zhou, K.~Yang, Y.~Shi, and K.~Wang, ``Reconfigurable intelligent surface for green edge inference,'' \emph{IEEE Trans. Green Commun. Netw.}, vol.~5, no.~2, pp. 964--979, 2021.

\bibitem{poularakis2019joint}
K.~Poularakis, J.~Llorca, A.~M. Tulino, I.~Taylor, and L.~Tassiulas, ``Joint service placement and request routing in multi-cell mobile edge computing networks,'' in \emph{IEEE INFOCOM 2019-IEEE Conf. Comput. Commun.}\hskip 1em plus 0.5em minus 0.4em\relax IEEE, 2019, pp. 10--18.

\bibitem{liang2021multi}
Z.~Liang, Y.~Liu, T.-M. Lok, and K.~Huang, ``Multi-cell mobile edge computing: Joint service migration and resource allocation,'' \emph{IEEE Trans. Wireless Commun.}, vol.~20, no.~9, pp. 5898--5912, 2021.

\bibitem{liang2022two}
------, ``A two-timescale approach to mobility management for multicell mobile edge computing,'' \emph{IEEE Trans. Wireless Commun.}, vol.~21, no.~12, pp. 10\,981--10\,995, 2022.

\bibitem{zhang2021intelligent}
S.~Zhang and R.~Zhang, ``Intelligent reflecting surface aided multi-user communication: Capacity region and deployment strategy,'' \emph{IEEE Trans. Commun.}, vol.~69, no.~9, pp. 5790--5806, 2021.

\bibitem{cai2021intelligent}
W.~Cai, R.~Liu, Y.~Liu, M.~Li, and Q.~Liu, ``Intelligent reflecting surface assisted multi-cell multi-band wireless networks,'' in \emph{2021 IEEE Wireless Commun. Netw. Conf. (WCNC)}.\hskip 1em plus 0.5em minus 0.4em\relax IEEE, 2021, pp. 1--6.

\bibitem{pan2020multicell}
C.~Pan, H.~Ren, K.~Wang, W.~Xu, M.~Elkashlan, A.~Nallanathan, and L.~Hanzo, ``Multicell {MIMO} communications relying on intelligent reflecting surfaces,'' \emph{IEEE Trans. Wireless Commun.}, vol.~19, no.~8, pp. 5218--5233, 2020.

\bibitem{zhang2021reconfigurable}
C.~Zhang, W.~Yi, Y.~Liu, K.~Yang, and Z.~Ding, ``Reconfigurable intelligent surfaces aided multi-cell {NOMA} networks: A stochastic geometry model,'' \emph{IEEE Trans. Commun.}, vol.~70, no.~2, pp. 951--966, 2021.

\bibitem{ni2021resource}
W.~Ni, X.~Liu, Y.~Liu, H.~Tian, and Y.~Chen, ``Resource allocation for multi-cell {IRS}-aided {NOMA} networks,'' \emph{IEEE Trans. Wireless Commun.}, vol.~20, no.~7, pp. 4253--4268, 2021.

\bibitem{hua2020intelligent}
M.~Hua, Q.~Wu, D.~W.~K. Ng, J.~Zhao, and L.~Yang, ``Intelligent reflecting surface-aided joint processing coordinated multipoint transmission,'' \emph{IEEE Trans. Commun.}, vol.~69, no.~3, pp. 1650--1665, 2020.

\bibitem{bai2021resource}
T.~Bai, C.~Pan, H.~Ren, Y.~Deng, M.~Elkashlan, and A.~Nallanathan, ``Resource allocation for intelligent reflecting surface aided wireless powered mobile edge computing in {OFDM} systems,'' \emph{IEEE Trans. Wireless Commun.}, vol.~20, no.~8, pp. 5389--5407, 2021.

\bibitem{yu2022irs}
J.~Yu, Y.~Li, X.~Liu, B.~Sun, Y.~Wu, and D.~H. Tsang, ``{IRS} assisted {NOMA} aided mobile edge computing with queue stability: Heterogeneous multi-agent reinforcement learning,'' \emph{IEEE Trans. on Wireless Commun.}, pp. 1--1, 2022.

\bibitem{xu2022deep}
J.~Xu, B.~Ai, L.~Chen, and L.~Wu, ``Deep reinforcement learning for communication and computing resource allocation in {RIS} aided {MEC} networks,'' in \emph{ICC 2022-IEEE International Conf. Commun.}\hskip 1em plus 0.5em minus 0.4em\relax IEEE, 2022, pp. 3184--3189.

\bibitem{rezaei2022energy}
A.~Rezaei, A.~Khalili, J.~Jalali, H.~Shafiei, and Q.~Wu, ``Energy-efficient resource allocation and antenna selection for {IRS}-assisted multi-cell downlink networks,'' \emph{IEEE Wireless Commun. Lett.}, vol.~11, no.~6, pp. 1229--1233, 2022.

\bibitem{wu2019towards}
Q.~Wu and R.~Zhang, ``Towards smart and reconfigurable environment: Intelligent reflecting surface aided wireless network,'' \emph{IEEE Commun. Mag.}, vol.~58, no.~1, pp. 106--112, 2019.

\bibitem{zheng2019intelligent}
B.~Zheng and R.~Zhang, ``Intelligent reflecting surface-enhanced {OFDM}: Channel estimation and reflection optimization,'' \emph{IEEE Wireless Commun. Lett.}, vol.~9, no.~4, pp. 518--522, 2019.

\bibitem{lu2021aerial}
H.~Lu, Y.~Zeng, S.~Jin, and R.~Zhang, ``Aerial intelligent reflecting surface: Joint placement and passive beamforming design with {3D} beam flattening,'' \emph{IEEE Trans. Wireless Commun.}, vol.~20, no.~7, pp. 4128--4143, 2021.

\bibitem{liu2020resource}
X.~Liu, J.~Yu, J.~Wang, and Y.~Gao, ``Resource allocation with edge computing in {IoT} networks via machine learning,'' \emph{IEEE Internet Things J.}, vol.~7, no.~4, pp. 3415--3426, 2020.

\bibitem{gurobi2018gurobi}
L.~Gurobi~Optimization, ``Gurobi optimizer reference manual,'' 2018.

\bibitem{smith1999symbolic}
E.~M. Smith and C.~C. Pantelides, ``A symbolic reformulation/spatial branch-and-bound algorithm for the global optimisation of nonconvex {MINLPs},'' \emph{Comput. Chem. Engineer.}, vol.~23, no. 4-5, pp. 457--478, 1999.

\bibitem{shen2018fractional}
K.~Shen and W.~Yu, ``Fractional programming for communication systems—part {I}: Power control and beamforming,'' \emph{IEEE Trans. Signal Process.}, vol.~66, no.~10, pp. 2616--2630, 2018.

\bibitem{sur2021sum}
S.~N. Sur, A.~K. Singh, D.~Kandar, and R.~Bera, ``Sum-rate analysis of intelligent reflecting surface aided multi-user millimeter wave communications system,'' in \emph{J. Phys.: Conf. Series}, vol. 1921, no.~1.\hskip 1em plus 0.5em minus 0.4em\relax IOP Publishing, 2021, p. 012050.

\bibitem{grant2014cvx}
M.~Grant and S.~Boyd, ``{CVX}: Matlab software for disciplined convex programming, version 2.1,'' 2014.

\bibitem{dampahalage2020intelligent}
D.~Dampahalage, K.~S. Manosha, N.~Rajatheva, and M.~Latva-aho, ``Intelligent reflecting surface aided vehicular communications,'' in \emph{GC Wkshps}.\hskip 1em plus 0.5em minus 0.4em\relax IEEE, 2020, pp. 1--6.

\end{thebibliography}

\begin{IEEEbiography}[{\includegraphics[width=1in,height=1.25in,clip,keepaspectratio]{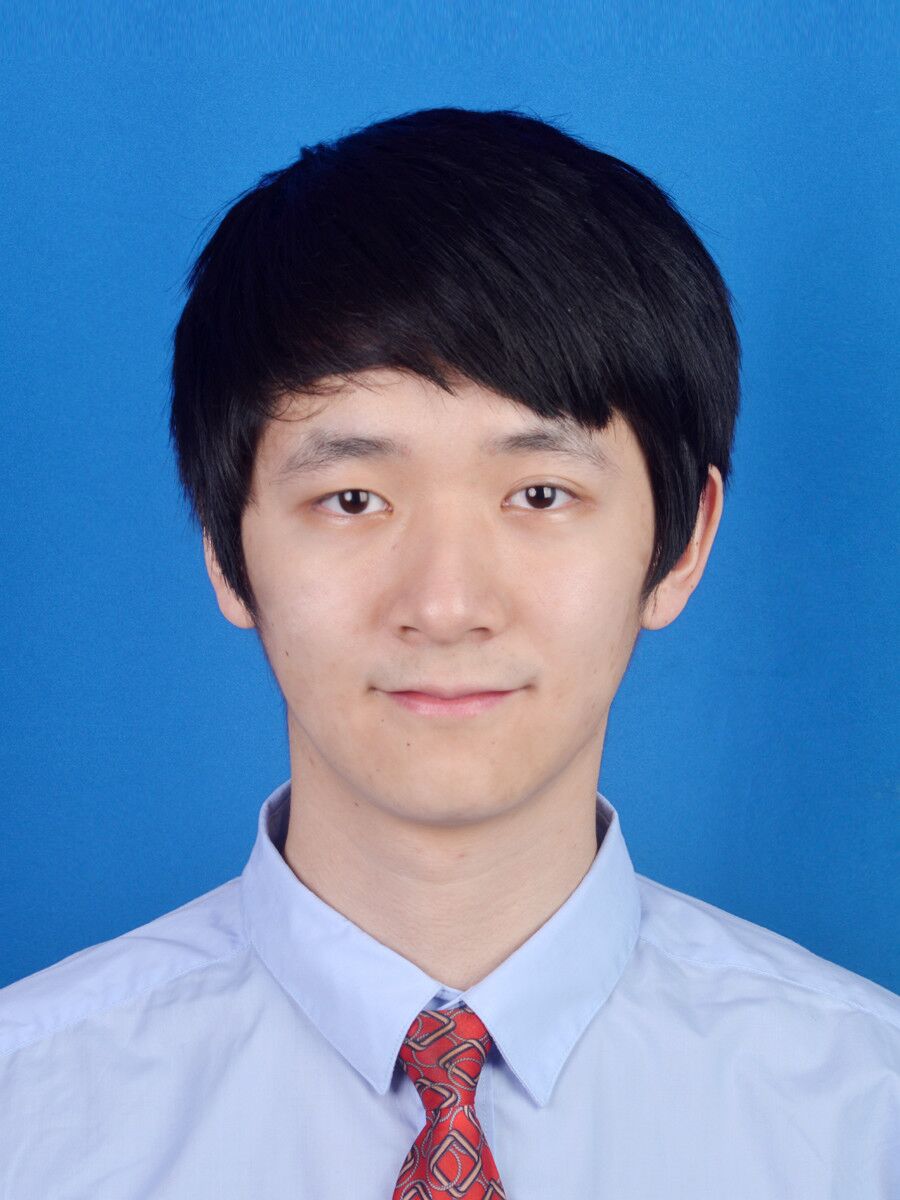}}]{Wenhan Xu} received the B.E. degree in information engineering from Southeast University, Nanjing, China, in 2020. He received his MPhil degree in electronic and computer engineering from the Hong Kong University of Science and Technology in 2023. He is currently a PhD student in Internet of Things Thrust, the Hong Kong University of Science and Technology, Guangzhou. His research interests are intelligent reflecting surface, mobile edge computing, and machine learning.
\end{IEEEbiography}

\begin{IEEEbiography}[{\includegraphics[width=1in,height=1.25in,clip,keepaspectratio]{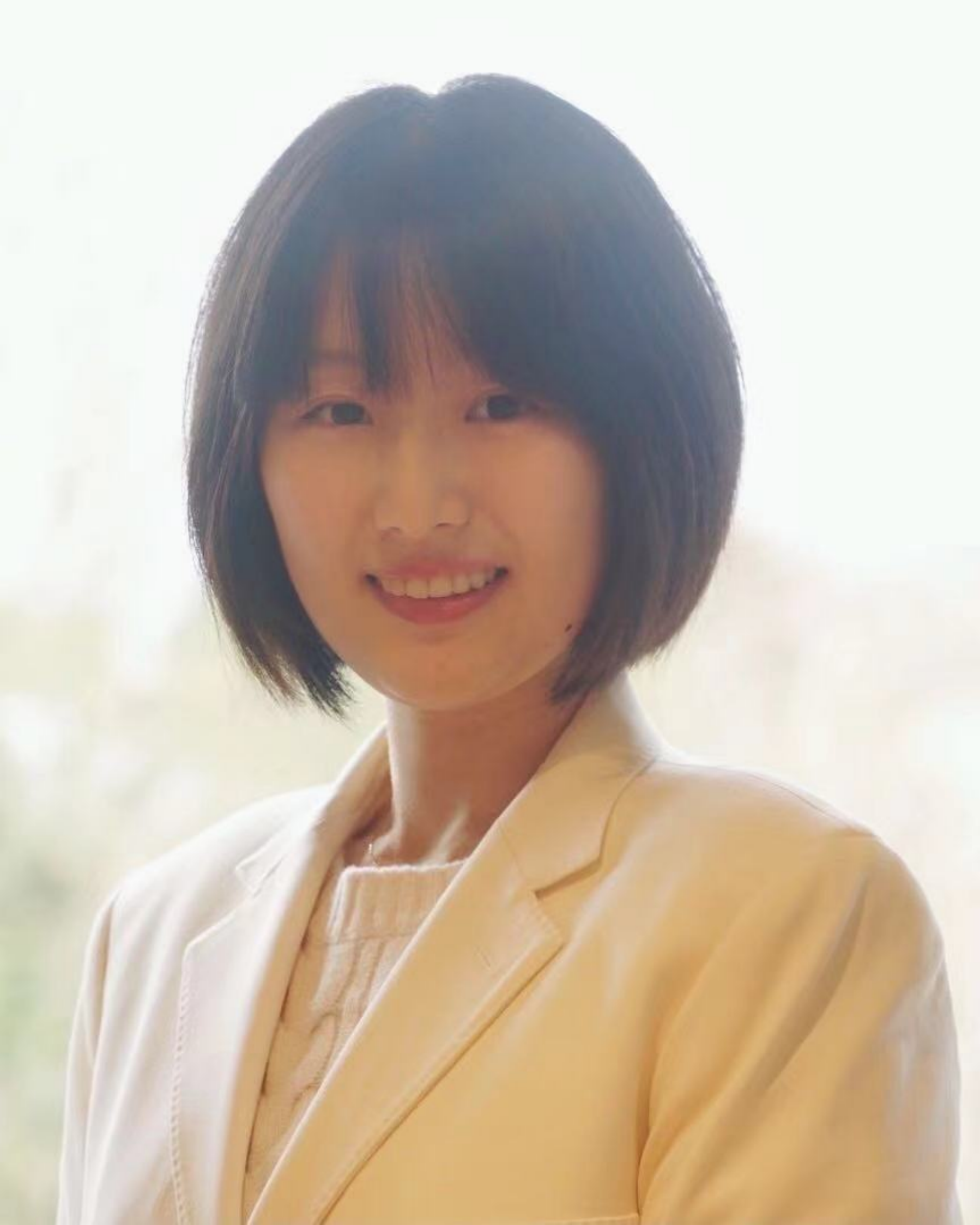}}]{Jiadong Yu} received the PhD degree from Queen Mary University of London, U.K. in 2021. She was a Teaching Fellow in Queen Mary University of London in 2021. Since Nov. 2021, she has been with Internet of Things Thrust, the Hong Kong University of Science and Technology, Guangzhou, where she is currently a Postdoctoral Fellow. Her current research interests include the machine learning, deep learning, and federated learning for internet of things and wireless communications.
\end{IEEEbiography}

\begin{IEEEbiography}[{\includegraphics[width=1in,height=1.25in,clip,keepaspectratio]{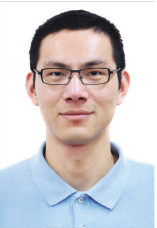}}]{Yuan Wu} (S'08-M'10-SM'16) is currently an Associate Professor with the State Key Laboratory of Internet of Things for Smart City, University of Macau, Macao, China, and also with the Department of Computer and Information Science, University of Macau. He received the PhD degree in Electronic and Computer Engineering from the Hong Kong University of Science and Technology in 2010. His research interests include resource management for wireless networks, green communications and computing, edge computing and edge intelligence, and energy informatics. He received the Best Paper Award from the IEEE ICC’2016, IEEE TCGCC’2017, IWCMC’2021, and IEEE WCNC’2023. Dr. Wu is currently on the editorial board of IEEE Transactions on Vehicular Technology, IEEE Transactions on Network Science and Engineering, and IEEE Internet of Things Journal.
\end{IEEEbiography}

\begin{IEEEbiography}[{\includegraphics[width=1in,height=1.25in,clip,keepaspectratio]{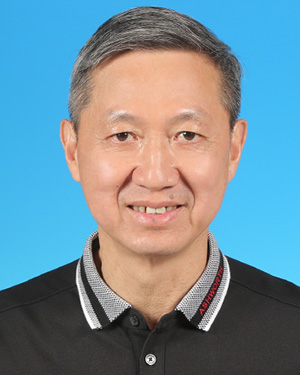}}]{Danny H.K. Tsang} received the Ph.D. degree in electrical engineering from the Moore School of Electrical Engineering, University of Pennsylvania, Philadelphia, PA, USA, in 1989. After graduation, he joined the Department of Computer Science at Dalhousie University, Halifax, NS, Canada. He later joined the Department of Electronic and Computer Engineering at the Hong Kong University of Science and Technology (HKUST), Hong Kong, in 1992, where he is currently a professor. He has also served as the Thrust Head of the Internet of Things Thrust at HKUST (Guangzhou), Guangzhou, China, since 2020. During his leave from HKUST from 2000 to 2001, he assumed the role of Principal Architect at Sycamore Networks, Chelmsford, MA, USA. His current research interests include cloud computing, edge computing, NOMA networks, and smart grids. Prof. Tsang was a Guest Editor of IEEE Journal on Selected Areas in Communications’ special issue on Advances in P2P Streaming Systems, an Associate Editor of Journal of Optical Networking published by the Optical Society of America, and a Guest Editor of IEEE Systems Journal. He currently serves as a member of the Special Editorial Cases Team (SECT) of IEEE Communications Magazine. He was responsible of the network architecture design of Ethernet MAN/WAN over SONET/DWDM networks. He invented the 64B/65B encoding (U.S. Patent No.: U.S. B2) and contributed it to the proposal for Transparent GFP in the T1X1.5 standard that was advanced to become the ITU G.GFP standard. The coding scheme has now been adopted by International Telecommunication Union (ITU)’s Generic Framing Procedure recommendation GFP-T (ITUT G.7041/Y.1303) and Interfaces of the Optical Transport Network (ITU-T G.709). He was nominated to become an IEEE Fellow in 2012 and an HKIE Fellow in 2013.
\end{IEEEbiography}

\end{document}